\documentclass[twocolumn]{aastex631}

\usepackage{multirow}
\usepackage{makecell}
\usepackage{comment}

\begin{document}

\title{Windy or not: Radio pc-scale evidence for a broad-line region wind in radio-quiet quasars}


\correspondingauthor{Sina Chen}
\email{sina.chen@campus.technion.ac.il}

\author{Sina Chen}
\affiliation{Physics Department, Technion, Haifa 32000, Israel}

\author{Ari Laor}
\affiliation{Physics Department, Technion, Haifa 32000, Israel}

\author{Ehud Behar}
\affiliation{Physics Department, Technion, Haifa 32000, Israel}

\author{Ranieri D. Baldi}
\affiliation{INAF - Istituto di Radioastronomia, via Gobetti 101, 40129 Bologna, Italy}

\author{Joseph D. Gelfand}
\affiliation{NYU Abu Dhabi, PO Box 129188, Abu Dhabi, UAE}

\author{Amy E. Kimball}
\affiliation{National Radio Astronomy Observatory, 1011 Lopezville Road, Socorro, NM 87801, USA}

\author{Ian M. McHardy}
\affiliation{Department of Physics and Astronomy, University of Southampton, Highfield, Southampton, SO17 1BJ, UK}

\author{Gabor Orosz}
\affiliation{Joint Institute for VLBI ERIC, Oude Hoogeveensedijk 4, 7991 PD Dwingeloo, The Netherlands}

\author{Zsolt Paragi}
\affiliation{Joint Institute for VLBI ERIC, Oude Hoogeveensedijk 4, 7991 PD Dwingeloo, The Netherlands}

\begin{abstract}

Does a broad-line region (BLR) wind in radio-quiet (RQ) active galactic nuclei (AGN) extend to pc scales and produce radio emission?
We explore the correlations between a pc-scale radio wind and the BLR wind in a sample of 19 RQ Palomar-Green (PG) quasars.
The radio wind is defined based on the spectral slope and the compactness of the emission at 1.5--5~GHz, and the BLR wind is defined by the excess blue wing in the C\,IV emission line profile.
The five objects with both radio and BLR wind indicators are found at high Eddington ratios $L/L_{\rm Edd}$ ($\ge$ 0.66), and eight of the nine objects with neither radio nor BLR winds reside at low $L/L_{\rm Edd}$ ($\le$ 0.28).
This suggests that the BLR wind and the radio wind in RQ AGN are related to a radiation pressure driven wind.
Evidence for free-free absorption by AGN photoionized gas, which flattens the spectral slope, is found in two objects.
Radio outflows in three low $L/L_{\rm Edd}$ (0.05--0.12) objects are likely from a low-power jet, as suggested by additional evidence.
The presence of a mild equatorial BLR wind in four intermediate $L/L_{\rm Edd}$ (0.2--0.4) objects can be tested with future spectropolarimetry.

\end{abstract}


\section{Introduction}

Active galactic nuclei (AGN) can be divided into Radio Loud (RL) and Radio Quiet (RQ) based on the radio 5~GHz to optical 4400~{\AA} flux ratio $R>10$ or $R<10$ \citep{Kellermann1989}.
The radio emission in RL AGN is produced by a powerful relativistic jet, with a characterized radio to X-ray luminosity ratio of $\log L_{\rm R}/L_{\rm X} \simeq -2$.
The origin of the radio emission in RQ AGN, characterized by $\log L_{\rm R}/L_{\rm X} \simeq -5$, is still under debate \citep{Laor2008}.
A variety of emission mechanisms are proposed, such as a low-power jet, an AGN-driven wind, the accretion disk corona, star formation (SF), and free-free emission \citep[see][for a review]{Panessa2019}.
How can we distinguish among the various possible radio emission mechanisms in RQ AGN?

A low-power jet and an AGN-driven wind could be similar in terms of the radio morphology and the spectral slope $\alpha$ using $S_{\nu} \propto \nu^{\alpha}$.
They both produce extended optically thin emission, which has a steep slope $\alpha < -0.5$ and may extend from mas to arcsec scales.
In principle, the jet is expected to be more collimated and propagate at a higher velocity than the wind. However, it is generally difficult to tell them apart based on radio observations.
Detailed studies of individual sources may allow to separate a jet from a wind.
For example, a proper motion study can reveal the bulk outflow velocity, and distinguish a relativistic jet from a non-relativistic wind.
In addition, a bulk relativistic speed would result in a significant Doppler effect, which strongly enhances the emission and leads to a high surface brightness temperature $T_{\rm B}$.

The accretion disk corona can produce compact optically thick emission, which is characterized by a flat or inverted slope $\alpha > -0.5$ \citep{Laor2008}.
Since it originates on sub-pc scales, it remains unresolved in the Very Long Baseline Interferometry (VLBI) observations, such as the Very Long Baseline Array (VLBA) and the European VLBI Network (EVN).
The corona can also produce extended optically thin emission, due to Coronal Mass Ejections (CME) in a form of outflowing plasma, as observed in the Sun and coronally active stars.
The accretion disk corona is also suggested to be a base for the formation of jet emission \citep{Merloni2002,King2017}.

SF produces diffuse optically thin emission, which has a steep slope $\alpha < -0.5$ and extends on the host galaxy scales \citep{Kimball2011,Condon2013,Zakamska2016}.
Free-free emission from AGN photoionized gas is characterized by a specific flat slope $\alpha = -0.1$, and it becomes self-absorbed with $\alpha = 2$ below a frequency set by the distance from the AGN \citep{Baskin2021}.
Both mechanisms produce emission with a low surface brightness temperature, typically well below $T_{\rm B} \sim 10^4 - 10^5$~K, which is resolved out in mas-scale VLBI observations.

In this paper we focus on looking for the pc-scale radio emission associated with an AGN-driven Broad-Line Region (BLR) wind, and how it can be distinguished from a low-power jet or the coronal emission.
We use the term ``core'' to refer to the unresolved radio emission, and the general term ``outflow'' to refer to both a wind and a jet when they are not specified.

An AGN-driven wind in RQ Quasars (RQQ), for example a radiation pressure driven wind \citep{Murray2005}, is expected to produce radio synchrotron emission through shocks and particle acceleration, which are generated by an interaction with the ambient medium \citep{Faucher-Giguere-2012}.
Outflows on kpc scales are indeed observed in radio imaging \citep{Gallimore2006,Smith2020} and in optical spectroscopy \citep{Zakamska2014,Rupke2017} in some RQQ, which may be evidence for such an interaction and a plausible feedback mechanism in regulating the growth of the supermassive black holes (BH) and their host galaxy \citep{Fabian2012}.

An earlier study based on archival Very Large Array (VLA) observations of 25 RQ Palomar-Green (PG) quasars \citep{Boroson1992} found a correlation between the 5--8.5~GHz spectral slope of the core emission ($<0.3$~arcsec) and the Eddington ratio \citep{Laor2019}.
All $L/L_{\rm Edd}>0.3$ objects have steep spectra ($\alpha_{5-8.5}<-0.5$), and nearly all objects with $L/L_{\rm Edd}<0.3$ have flat spectra ($\alpha_{5-8.5}>-0.5$).
This trend is further confirmed by a VLA study of RQ PG quasars at 45~GHz \citep{Baldi2022}.
Since a steep spectral slope implies optically thin emission from an extended source, and high $L/L_{\rm Edd}$ is likely associated with a radiatively driven wind \citep[e.g.][]{Baskin2005}, the steep radio emission of high $L/L_{\rm Edd}$ RQQ may originate from shocks generated by an AGN wind.

Follow-up VLBA studies of 18 RQ PG quasars found an interesting correlation where the 1.5--5~GHz spectral slopes increase with the increasing core to total flux ratios, that is the source becomes more compact as the spectral slope becomes flatter \citep{Alhosani2022,Chen2023}.
This indicates that the radio emission in the steep-spectrum objects is more extended than that in the flat-spectrum objects, and the extended emission is likely produced by an AGN-driven wind, or possibly a low-power jet.
Thus the spectral slope and the emission compactness provide important hints on the origin of the radio emission.

The winds can be observed through absorption lines, when they are directly along our line of sight.
Ultraviolet (UV) and X-ray absorbing winds are commonly observed in AGN through UV and X-ray spectroscopy, and show an increasing maximal wind velocity with UV luminosity \citep{Laor2002,Ganguly2008,Gibson2009}, from $\sim$ 1,000~km\,s$^{-1}$ in Seyfert galaxies to $\sim$ 10,000~km\,s$^{-1}$ in broad absorption line quasars \citep[BALQ;][]{Reichard2003}.
The UV and X-ray absorption in BALQ are usually correlated, which suggests that high-velocity winds are often present in objects with strong UV absorption and weak X-ray emission \citep{Brandt2000,Laor2002,Gibson2009}.
Sometimes, UV and X-ray absorption variability matches are found \citep{Kaastra2014,Kriss2019,Mehdipour2022}, which indicates that they are caused by the same outflowing material.

High ionization UV emission lines, specifically the C\,IV line profile, which often shows blueshift and excess blue wing emission, are also an indicator of the winds from the BLR \citep{Richards2002,Richards2011}.
The advantage of emission lines is that the wind does not need to be directed along the line of sight (as in absorption lines), and the wind emission can be viewed from a wider angles.
In the PG quasar sample of 87 $z<0.5$ bright AGN, of which 71 are RQ \citep{Boroson1992}, a fair fraction (about one fifth) of the objects indeed show excess blue wing emission in the C\,IV emission line in the Hubble Space Telescope (HST) or International Ultraviolet Explorer (IUE) UV spectroscopy \citep{Baskin2005}, which allows to study the winds in a larger AGN population.
Such low-$z$ RQ objects are more suitable for follow-up mas-scale radio observations in order to resolve the nuclear winds than high-$z$ objects.

The winds indicated by the UV emission or absorption lines, are likely produced on the BLR scales \citep{Hamann1993}, which has a typical size increasing from 0.01~pc at Seyfert luminosity to 0.1~pc at quasar luminosity \citep{Kaspi2000}.
They may extend to the host galaxy $\sim$ kpc scales and lead to the AGN feedback \citep{Tombesi2015}.
Do these nuclear winds extend significantly outside the BLR, or do they form a failed wind \citep{Czerny2011}?
High resolution VLBI imaging of the radio emission would help to resolve the spatial scales of these nuclear winds.

Radio VLBI observations indeed provide evidence that the pc-scale radio emission is possibly associated with the BLR winds.
A specific example is PG\,1700+518 which is a RQ BALQ \citep{Blundell1998}.
An EVN observation of this object reveals a compact core and two small close to symmetric lobes about 1~kpc from the core \citep{Yang2012}.
Another example is PG\,0003+199 (Mrk\,335), a RQQ ($\log L_{\rm R}/L_{\rm X} = -5.44$) characterized by a steep radio slope \citep[$\alpha_{5-8.5} = -0.86$;][]{Laor2019} and high Eddington ratio ($L/L_{\rm Edd} \simeq 1$).
It shows C\,IV excess blue wing emission \citep{Baskin2005} suggestive of a BLR wind, and its VLBA image shows an elongated structure extending $\sim 10$~pc to the south of the optical position, and a somewhat fainter and more diffuse structure extending to the north \citep{Yao2021}.

Is the pc-scale radio emission produced by a wide-angle wind, as indicated by the broad emission or absorption line, or by a collimated jet, with the luminosity of about a factor of $\sim 10^3$ weaker than the one in RL Quasars (RLQ)?
Is such pc-scale structure representative of an AGN BLR wind? Do AGN without a BLR wind lack this emission pattern?
These open questions are the motivation of our study.

In this work, we use a sample of 19 PG RQQ to study the correlation between the radio pc-scale emission in the L (1.5~GHz) and C (5~GHz) bands combining EVN and VLBA observations, and the BLR wind indicated by the excess blue wing in the C\,IV emission line.
Out of the 19 objects, six are from our new EVN observations (10 observed) analyzed here, and 13 are from our earlier VLBA studies \citep[18 observed;][]{Alhosani2022,Chen2023}.
We aim to address the following questions.
(a) Is the pc-scale radio emission in RQQ related to the C\,IV excess blue wing or the BLR wind?
(b) Do objects with the BLR wind show extended radio emission on pc scales?
(c) Do objects without the BLR wind show only unresolved radio emission?
If the pc-scale radio emission is produced by an AGN-driven nuclear wind, we expect a relation between the pc-scale radio emission and the other AGN-driven nuclear wind indicators. The search for such a relation is the purpose of this study.

The paper is organized as follows. In Section 2 we describe the sample selection, in Section 3 we present the EVN observation and data reduction, in Section 4 we describe the EVN data analysis methods, and in Section 5 we present the classification of the radio emission in the EVN observations. In section 6 we describe the correlations between the radio and the BLR winds in a combined EVN and VLBA sample, and discuss them in section 7. Section 8 provides a summary of the main results.

\section{Sample selection}

The PG quasar sample \citep{Boroson1992} comprises $\sim$ 100 of the brightest AGN, which are selected based on a point-like morphology, blue colors, and the presence of broad emission lines \citep{Schmidt1983}.
These criteria produce a homogeneous and representative sample of Type 1 AGN, which are mostly at high $L/L_{\rm Edd}$ (0.1--1) and not significantly reddened.
This is likely the most extensively studied samples of Type 1 AGN, including: the overall Spectral Energy Distribution (SED) \citep{Neugebauer1987,Sanders1989}, radio cm-band continuum and imaging \citep{Kellermann1989,Kellermann1994,Miller1993}, infrared photometry \citep{Haas2003,Shi2014,Petric2015}, optical spectroscopy \citep{Boroson1992}, optical polarization \citep{Berriman1990}, UV spectroscopy \citep{Baskin2005}, soft X-ray spectroscopy \citep{Brandt2000}, and many other studies.
A critical property of the PG quasar sample is that it is optically selected, and therefore not subject to a selection bias in terms of its radio properties.

The sample is selected from the 71 RQQ among the 87 $z<0.5$ PG quasars.
We only considered 38 objects which were detected at 5~GHz with the VLA A-configuration \citep{Kellermann1989}, and further excluded 18 objects that had been observed with the VLBA in the L and C bands in our previous studies \citep{Alhosani2022,Chen2023}.
Of the remaining 20 objects, we selected five objects which show the most evident excess blue wing or blueshift in the C\,IV emission line, and another five objects which have the most symmetric C\,IV emission line as a control sample, based on an eye inspection of the C\,IV profile compared to the H$\beta$ profile in the UV spectra \citep{Baskin2005}.
Both line profiles are of the BLR emission only, as any possible contribution from the narrow-line region (NLR) is subtracted using the observed [O\,III]$\lambda$5007 line profile.
To improve the statistics, we later include in the analysis 13 of the 18 RQ PG quasars, which were detected with the VLBA in the L and C bands in our previous studies \citep{Alhosani2022,Chen2023}, and also have C\,IV emission line in the UV spectra \citep{Baskin2005}.

\begin{figure*}[ht!]
\centering
\includegraphics[width=1.7\columnwidth, trim={0cm, 0cm, 0cm, 0cm}, clip]{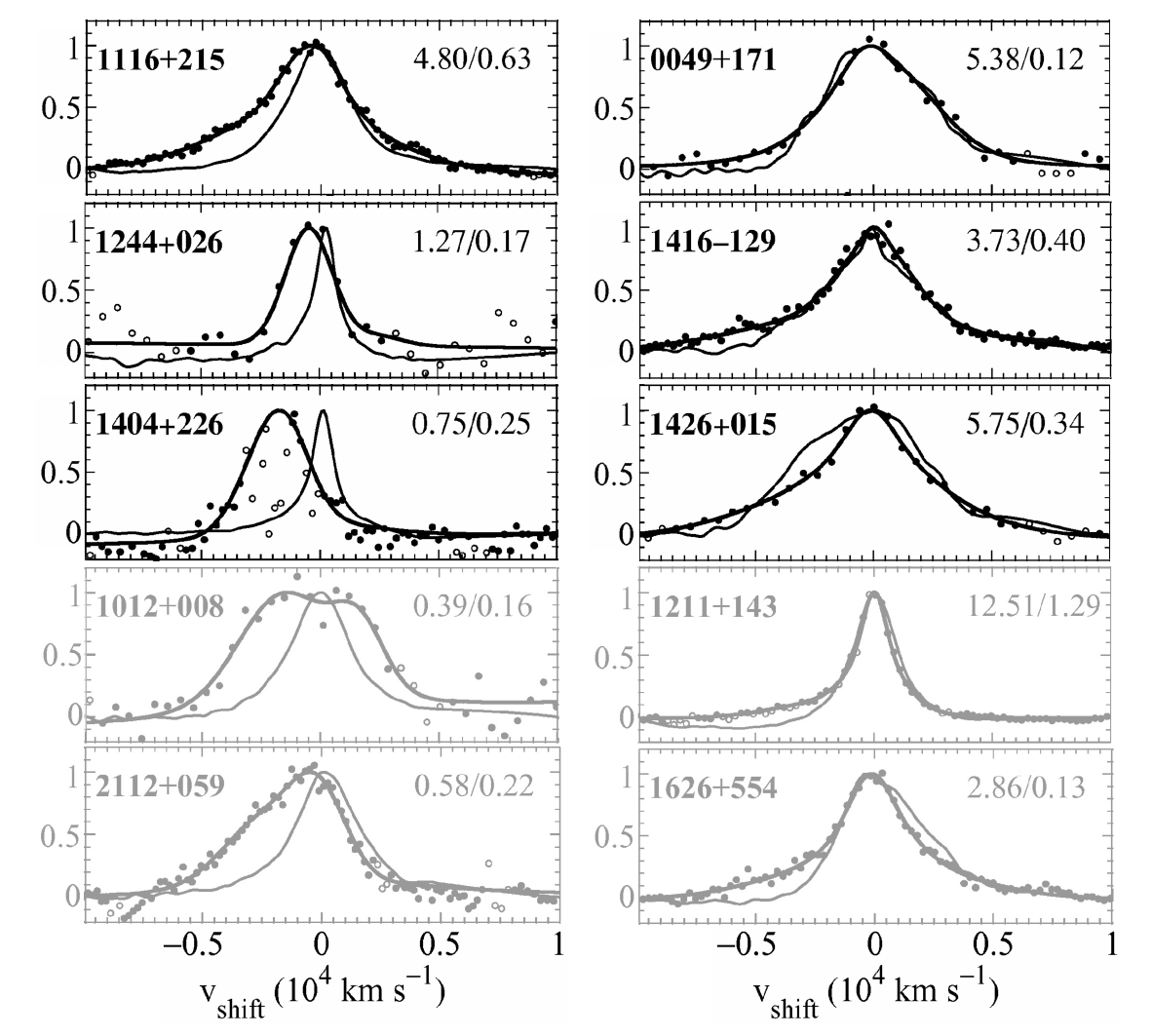}
\caption{The C\,IV emission line profile compared to the H$\beta$ line profile of the objects observed with the EVN, adapted from Fig.\,1 in \citet{Baskin2005}. Both line profiles are of the net BLR emission. The objects in the 1$^{st}$ column are selected to show C\,IV blue excess, and the objects in the 2$^{nd}$ column do not show C\,IV blue excess. The 6 objects in the 1$^{st}$, 2$^{nd}$, and 3$^{rd}$ rows are detected in the EVN observations, and the 4 objects in the 4$^{th}$ and 5$^{th}$ rows are not detected in the EVN observations.
The name of the object is indicated in the top-left corner of each panel.
The thick line is the C\,IV profile and the thin line is the H$\beta$ profile. Both profiles are normalized by their peak flux density.
The peak flux density of C\,IV/H$\beta$ lines is listed in the top-right corner of each panel in units of $10^{-14}$~erg\,cm$^{-2}$\,s$^{-1}$\,{\AA}$^{-1}$.
The filled circles are data points used in the fitting procedure, and the empty circles are data points excluded due to possible intrinsic or Galactic absorption.}
\label{civ_evn}
\end{figure*}

\begin{figure*}[ht!]
\centering
\includegraphics[width=1.8\columnwidth, trim={0cm, 0cm, 0cm, 0cm}, clip]{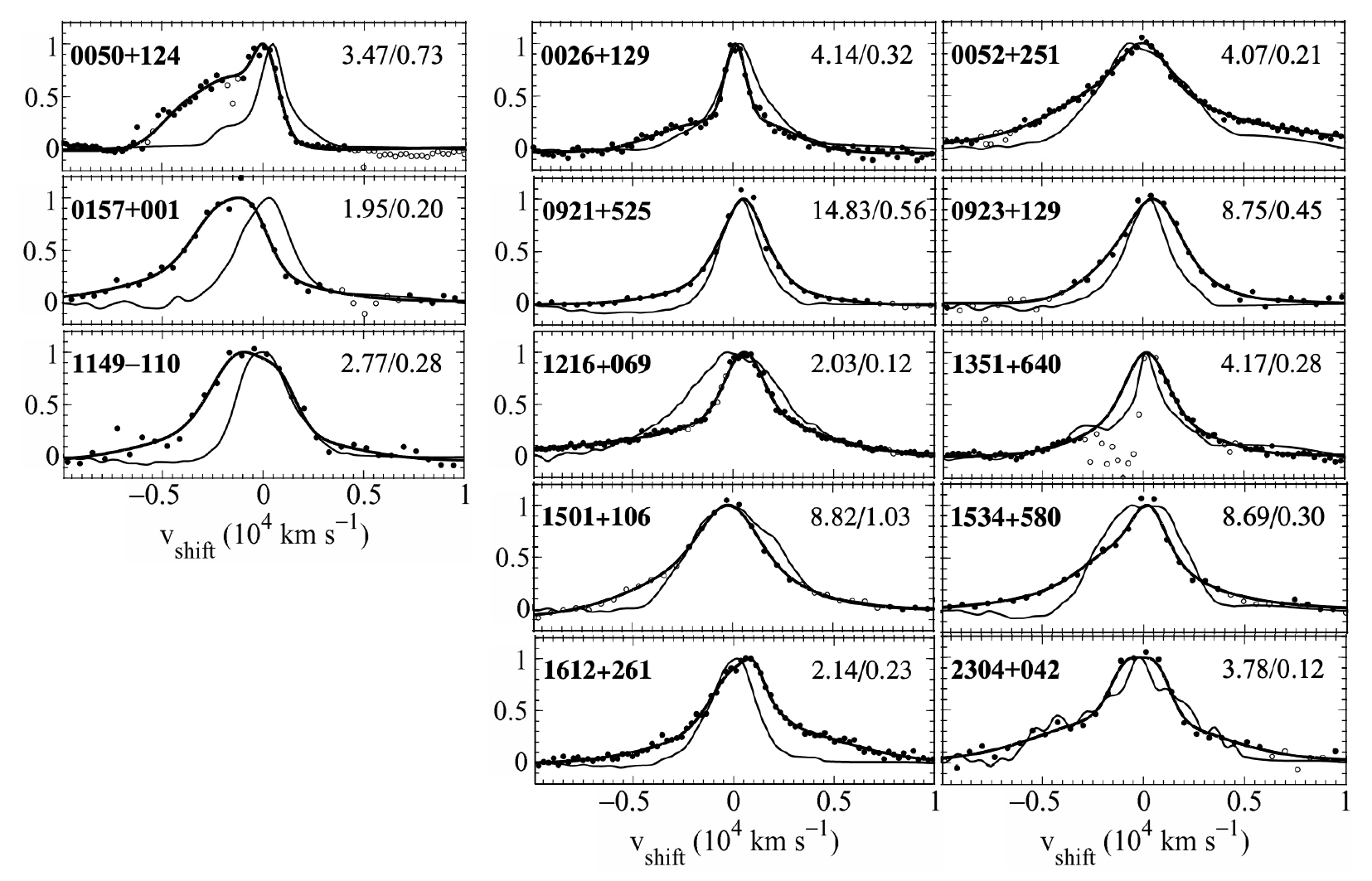}
\caption{The C\,IV emission line profile compared to the H$\beta$ line profile of the objects detected with the VLBA, adapted from Fig.\,1 in \citet{Baskin2005}. The 1$^{st}$ column shows the 3 objects with C\,IV blue excess, and the 2$^{nd}$ and 3$^{rd}$ columns show the 10 objects without C\,IV blue excess. The labels are the same as Figure~\ref{civ_evn}.}
\label{civ_vlba}
\end{figure*}

The C\,IV profiles compared to the H$\beta$ profiles of the 10 objects observed with the EVN are plotted in Figure~\ref{civ_evn}, including five objects with the C\,IV emission line showing strong excess blue wing or blueshift, and five objects with the C\,IV emission line close to symmetric.
The C\,IV profiles of the additional 13 objects detected with the VLBA are presented in Figure~\ref{civ_vlba}, including three objects with a significant C\,IV blue excess and ten objects without.
In the combined EVN and VLBA sample, the ratio of the C\,IV velocity shift to the full width at half maximum (FWHM) is between $-$0.538 and $-$0.040 in the objects with a C\,IV blue excess, and between $-$0.058 and 0.180 in the objects without a C\,IV blue excess.

\begin{figure}[ht!]
\centering
\includegraphics[width=\columnwidth, trim={0cm, 0cm, 1cm, 1cm}, clip]{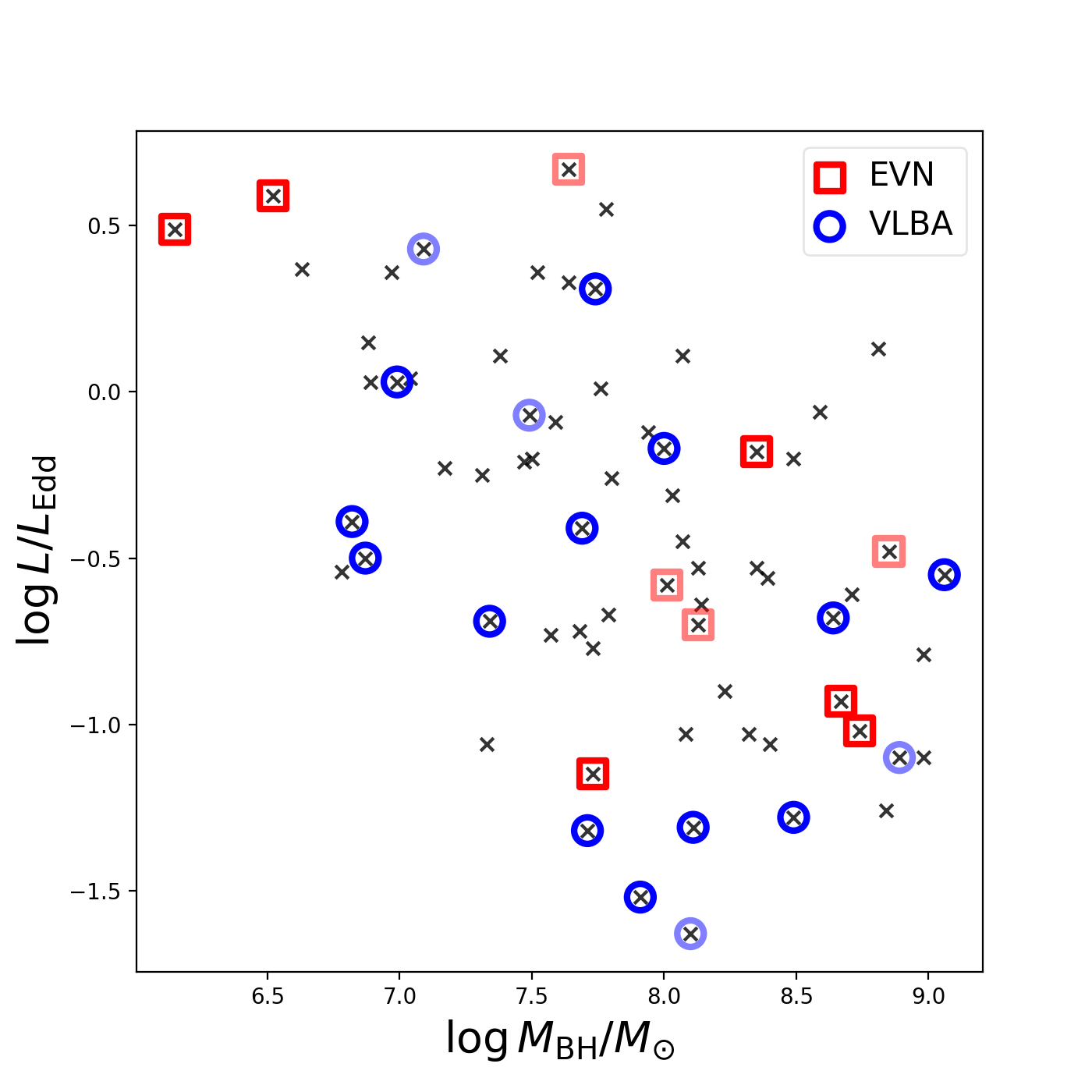}
\caption{The distribution of the BH mass and the Eddington ratio of the combined EVN and VLBA sample. The black crosses represent the 71 $z<0.5$ RQ PG quasars. The red squares and the blue circles mark the objects observed with the EVN and the VLBA respectively. The undetected objects are in fainter colors.}
\label{sample}
\end{figure}

The BH mass and the Eddington ratio in the sample cover a wide range, with $\log M_{\rm BH}/M_{\odot}$ from 6.2 to 9.1 and $\log L/L_{\rm Edd}$ from $-$1.5 to 0.6.
Figure~\ref{sample} shows the distribution of $\log M_{\rm BH}/M_{\odot}$ and $\log L/L_{\rm Edd}$ of the combined EVN and VLBA sample, which is consistent with that of the parent sample of 71 $z<0.5$ PG RQQ, and is likely representative of the general properties of the parent sample.

\section{Observations and data reduction}

The EVN observations (Program ID: EC088), including enhanced Multi Element Remotely Linked Interferometer Network (e-MERLIN), were carried out between 2022 October and 2023 March in the L and C bands, centered at 1.7 and 4.9~GHz respectively.
The data was recorded at a rate of 2048 Mbps in the C band and 1024 Mbps in the L band at the EVN stations (Jb, Wb, Ef, Mc, Nt, O8, Tr, Ys, Hh, Ir, T6, Ur, Km), and 512 Mbps at the e-MERLIN stations (Cm, Da, De, Kn, Pi) in both bands.
The data correlation was done by the EVN software correlator \citep[SFXC;][]{Keimpema2015} at the Joint Institute for VLBI ERIC (JIVE) using standard correlation parameters of continuum experiments (32~MHz dual-polarization subbands with 64 channels each).
The objects 3C454.3 and 4C39.25 were used as a fringe finder and bandpass calibrator.
All the observations were carried out in the phase-referencing mode with a switching cycle of 5 minutes.
The total observation time is two hours for each object in each band, and about 60\% of the total time is on target.

The visibility data was calibrated with the NRAO Astronomical Image Processing System
\citep[AIPS;][]{Greisen2003} following the EVN data reduction guide \footnote{\url{https://www.evlbi.org/evn-data-reduction-guide}}.
The standard steps include: \\
(i) Initial amplitude calibration using system temperatures and gain curves from each antenna, or nominal system equivalent flux densities in case of missing these data. \\
(ii) The ionospheric dispersive delays were corrected according to maps of total electron content provided by Global Positioning System satellite observations. \\
(iii) The phase errors due to instrumental delays at each antenna were removed using the fringe finder. \\
(iv) The bandpass calibration was performed using the fringe finder. \\
(v) The frequency and time-dependent phase calibration was performed for the whole observation using a nearby phase calibrator.

The phase calibrator imaging and self-calibration procedures were performed in DIFMAP \citep{Shepherd1994} through a number of iterations of model fitting with a point source function.
We then re-ran the fringe fitting and the amplitude and phase self-calibration on the phase calibrator in AIPS with the input source model made in DIFMAP.
The final solutions on the phase calibrator were transferred to the targets via linear interpolation.

The images of the targets were also produced in DIFMAP.
Inspection on all baselines (pairs of antennas) and spectral windows for radio frequency interference (RFI) was performed on the phase calibrator.
Data suffered from RFI were flagged on both the phase calibrator and the target.
We chose natural weighting which maximizes sensitivity at the expense of angular resolution.
A self-calibration was not applied on the targets since the signal-to-noise (S/N) ratio is not high enough.
In order to measure the spectral slope, which is less biased by the resolution in the different bands, we tapered the images with the same $uv$-range in both L and C bands.

The final images were inspected using the AIPS task IMEAN to obtain the background noise in a source-free region.
The AIPS task JMFIT was used to model the source with a 2D Gaussian profile, to obtain the peak intensity, the integrated flux density, the source position, and the source sizes before and after deconvolution.
We leave the centroid location, peak intensity, major and minor axes, and position angle, as free parameters in the Gaussian model.
In a few observations, the Effelsberg station, which is the most sensitive antenna in the array and usually used as a reference, was out of observations due to heavy snow, which resulted in a higher noise level than at the other epochs.

\section{Data analysis}

\setcounter{figure}{3}
\begin{figure*}
\gridline{\fig{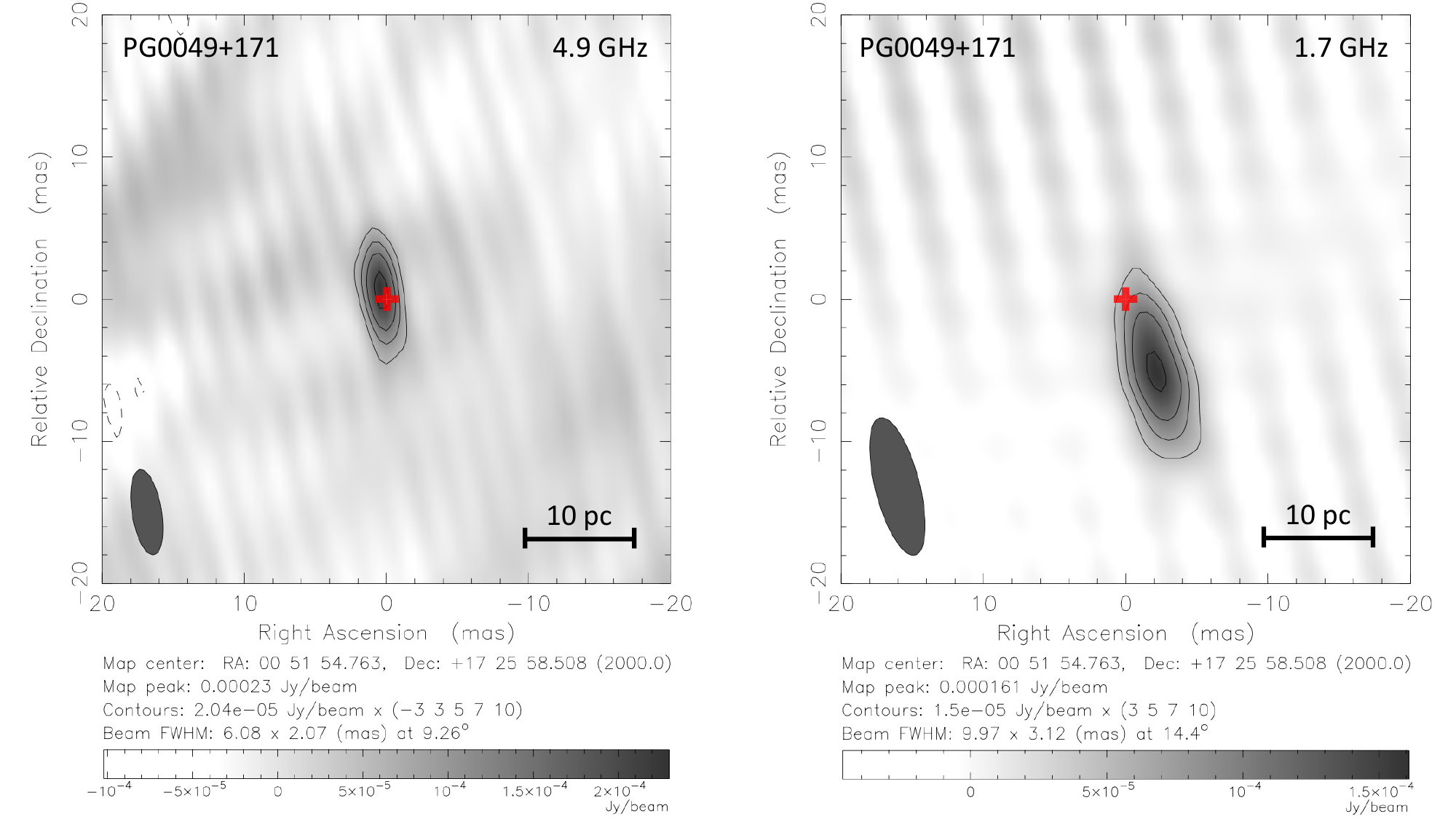}{.9\textwidth}{(a) PG\,0049+171: The contours are at ($-$3, 3, 5, 7, 10) $\times$ 0.0204~mJy/beam at 4.9~GHz (left) and (3, 5, 7, 10) $\times$ 0.0150~mJy/beam at 1.7~GHz (right).}}
\gridline{\fig{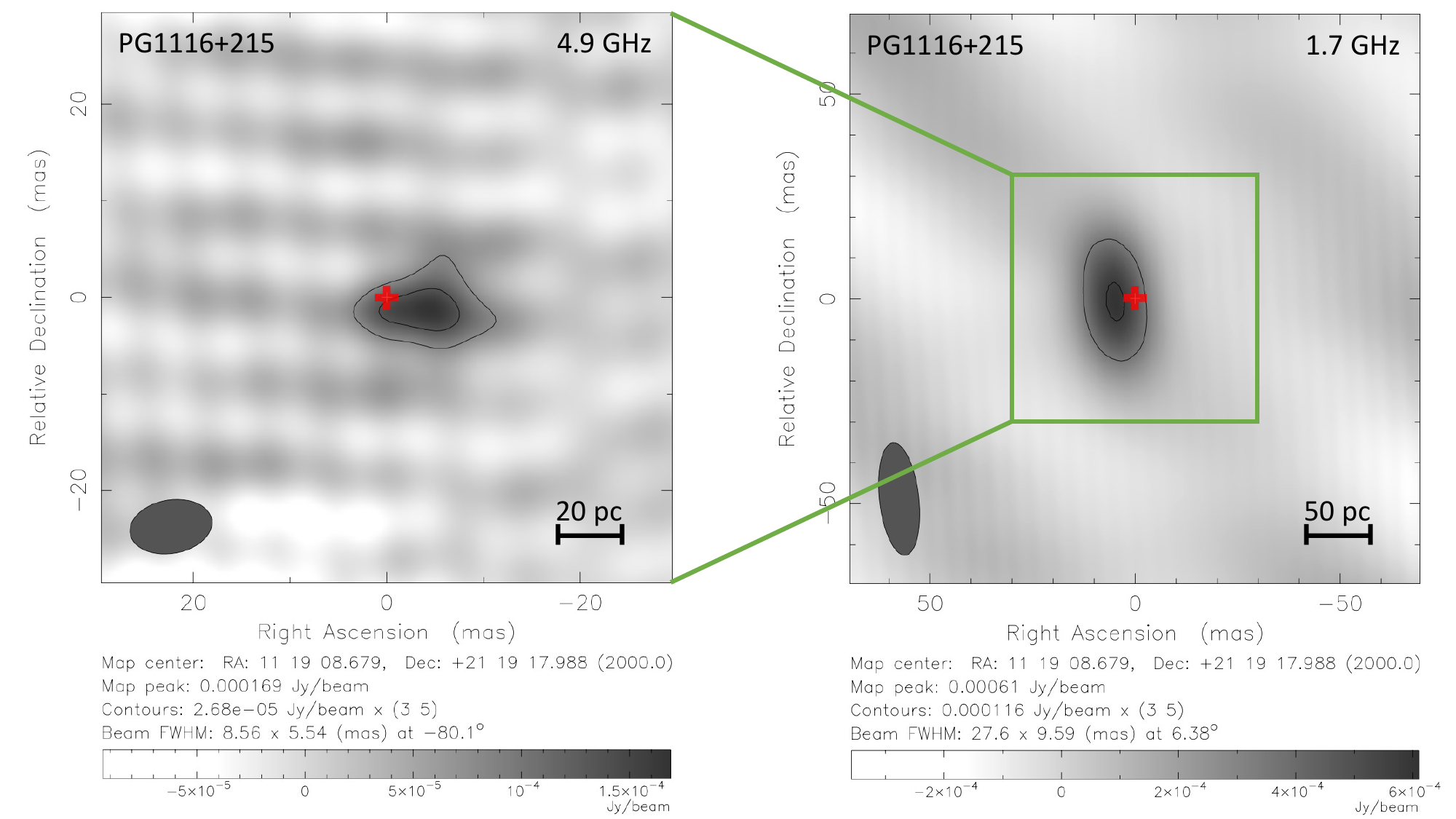}{.9\textwidth}{(b) PG\,1116+215: The tapered maps with the same $uv$-range at both frequencies are shown. The contours are at (3, 5) $\times$ 0.0268~mJy/beam at 4.9~GHz (left) and (3, 5) $\times$ 0.116~mJy/beam at 1.7~GHz (right).}}
\caption{Radio maps of the 6 objects detected in the EVN observations at 1.7 and 4.9~GHz. The synthesized beam is shown in the lower-left corner, and the size and orientation are reported in Table~\ref{size}. The $uv$-range and the background noise RMS are reported in Table~\ref{flux}. The images are centered at the {\it Gaia} position, which is marked as a red plus. The name of the object and the central frequency are indicated in the top-left and top-right corners of each panel respectively.}
\label{maps}
\end{figure*}

\setcounter{figure}{3}
\begin{figure*}
\gridline{\fig{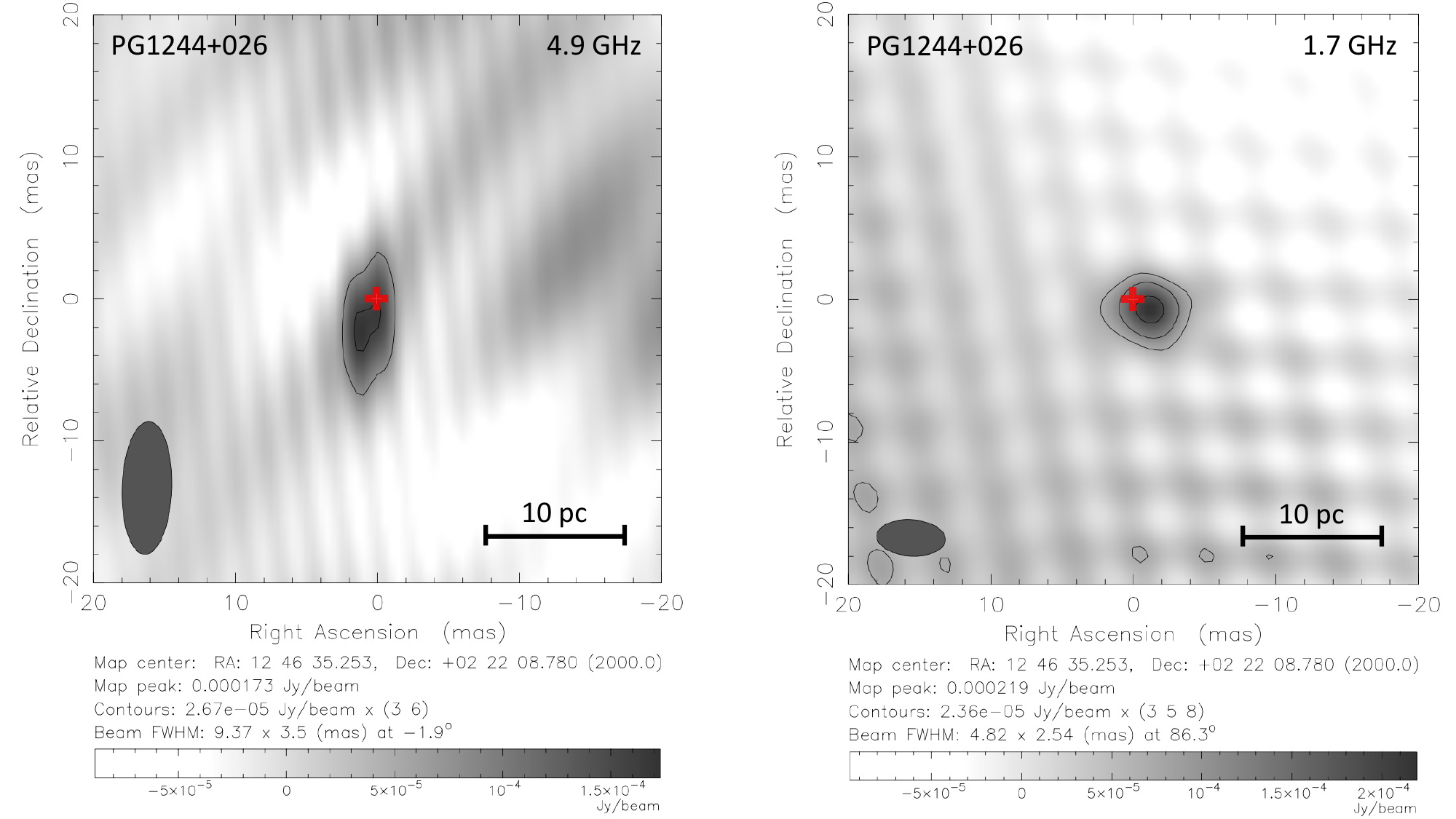}{.9\textwidth}{(c) PG\,1244+026: The contours are at (3, 6) $\times$ 0.0267~mJy/beam at 4.9~GHz (left) and (3, 5, 8) $\times$ 0.0236~mJy/beam at 1.7~GHz (right).}}
\gridline{\fig{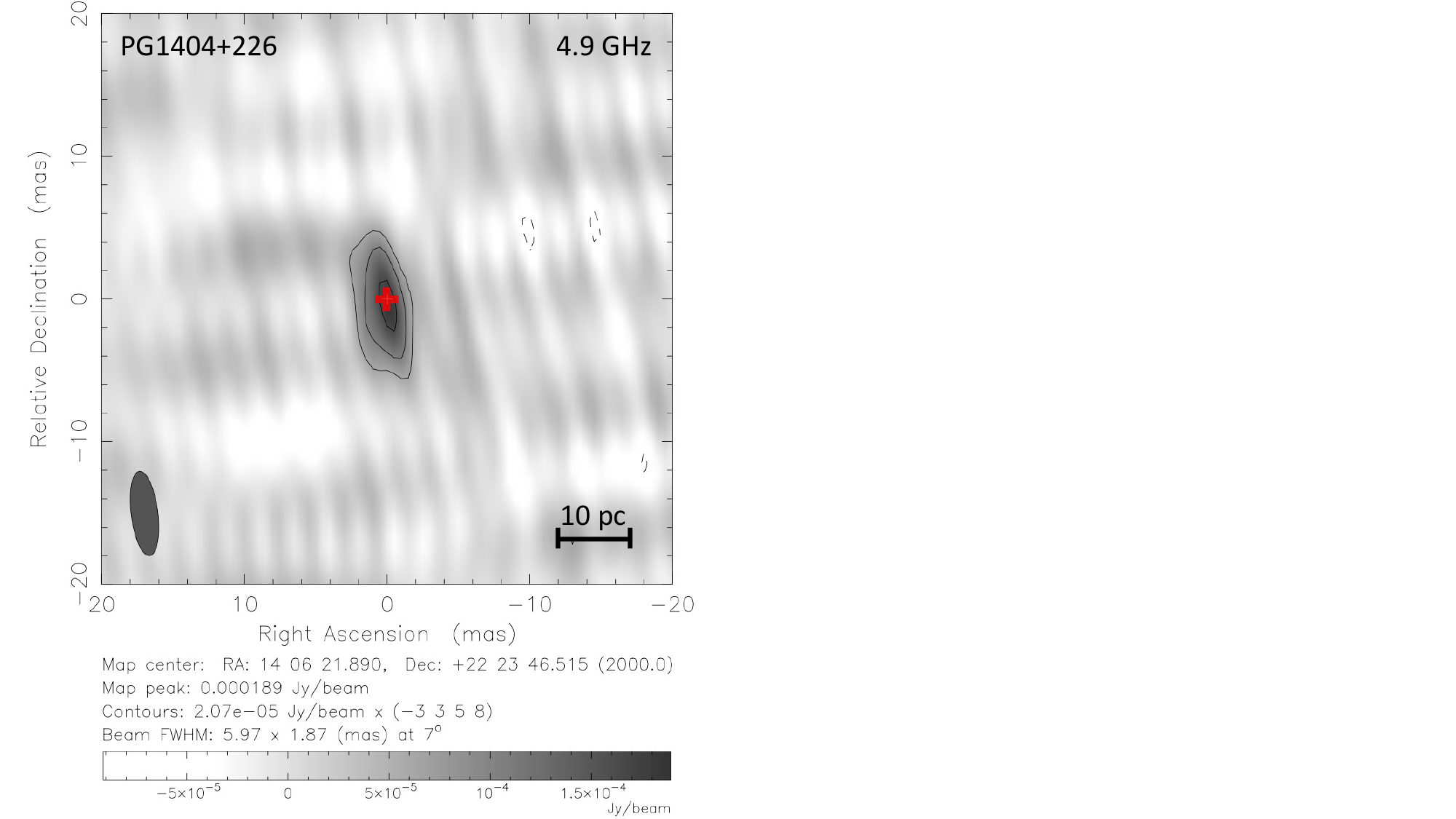}{.9\textwidth}{(d) PG\,1404+226: The contours are at ($-$3, 3, 5, 8) $\times$ 0.0207~mJy/beam at 4.9~GHz. The object is not detected at 1.7~GHz.}}
\caption{Continued.}
\end{figure*}

\setcounter{figure}{3}
\begin{figure*}
\gridline{\fig{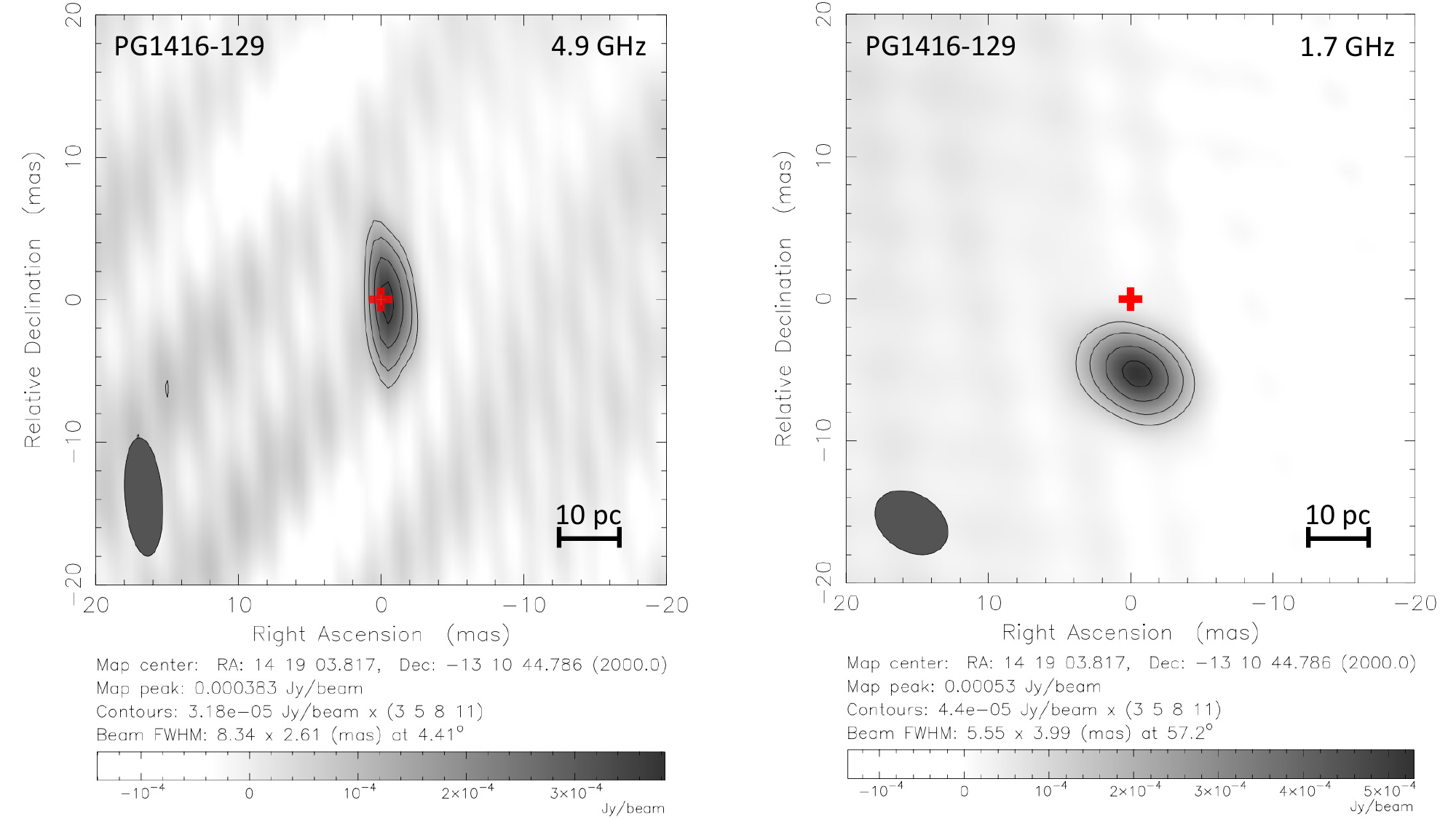}{.9\textwidth}{(e) PG\,1416$-$129: The contours are at (3, 5, 8, 11) $\times$ 0.0318~mJy/beam at 4.9~GHz (left) and (3, 5, 8, 11) $\times$ 0.0440~mJy/beam at 1.7~GHz (right).}}
\gridline{\fig{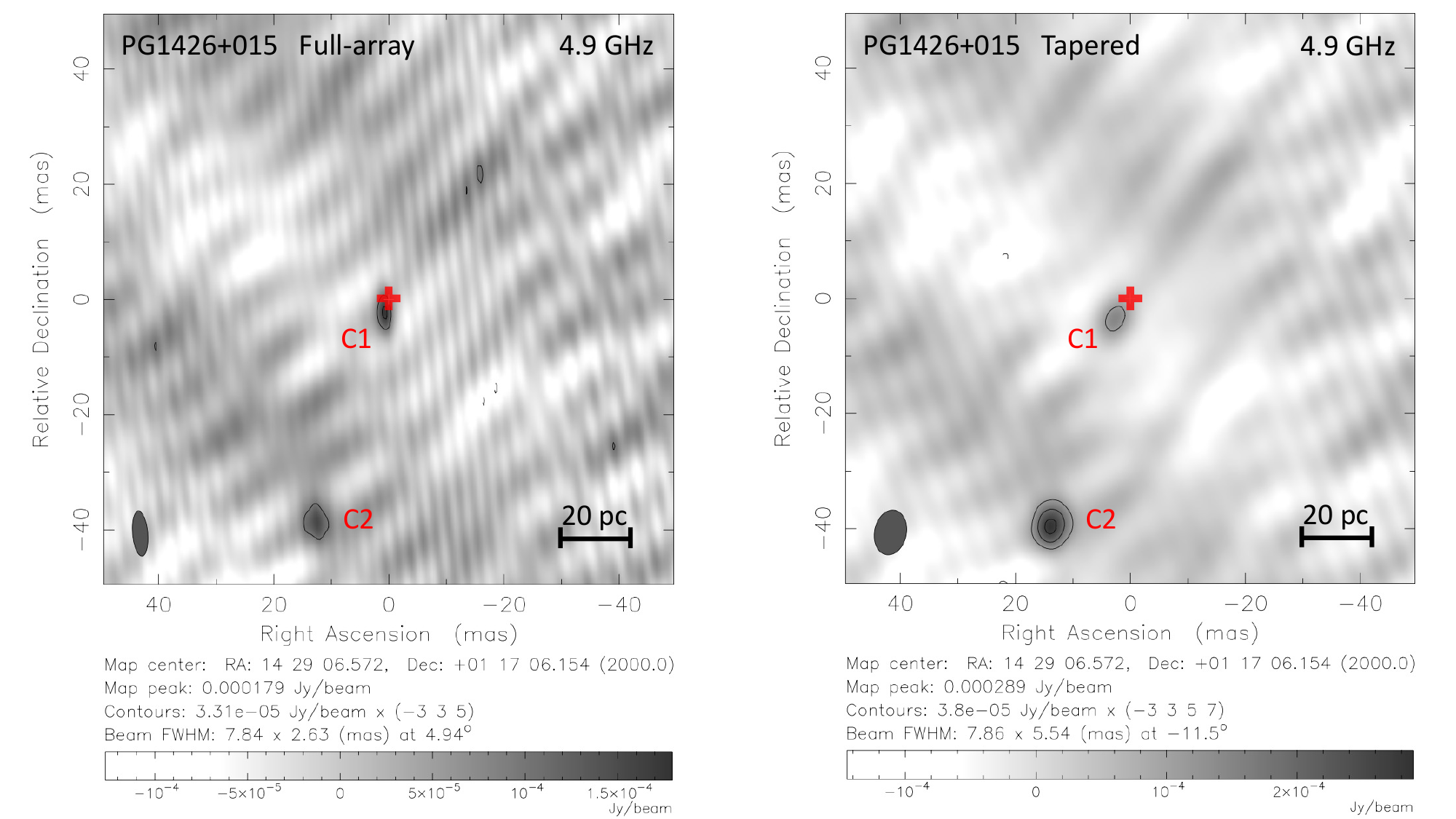}{.9\textwidth}{(f) PG\,1426+015: The contours are at ($-$3, 3, 5) $\times$ 0.0331~mJy/beam in the full-array map (left) and ($-$3, 3, 5, 7) $\times$ 0.0380~mJy/beam in the tapered map (right). Both images are at 4.9~GHz, the source is not detected at 1.7~GHz.}}
\caption{Continued.}
\end{figure*}

We consider a 5$\sigma$ detection as the detected criterion, where $\sigma$ is the background noise.
Figure~\ref{maps} presents the radio maps of the 6 RQ PG quasars detected in the EVN observations at 1.7 and 4.9~GHz centered at the {\it Gaia} position \citep{Gaia2016,Gaia2023}.
PG\,0049+171, PG\,1244+026, and PG\,1416$-$129 are detected at both frequencies.
PG\,1116+215 is detected only after tapering to the same $uv$-range at both frequencies, and PG\,1404+226 and PG\,1426+015 are detected only at 4.9~GHz.
The full-array and tapered images of PG\,1426+015 at 4.9~GHz are both presented.
It has two components, and it is the only source with this radio morphology in our EVN sample, but there are others in our VLBA sample.
The core component (C1) is detected at a 5$\sigma$ level in the full-array map, but at $< 5\sigma$ in the tapered map due to a higher noise level.
In contrast, the extended component (C2) is detected only at 3$\sigma$ in the full-array map, but at 7$\sigma$ in the tapered map.

Table~\ref{size} lists the EVN positions and their distances from the {\it Gaia} positions \citep{Gaia2016,Gaia2023}.
The offsets between the {\it Gaia} and the VLBI positions are found to be in a range of $\sim$ 0.1--10~mas in about 90\% of the AGN population with a median value of $\sim$ 2~mas \citep{Petrov2017}.
Both the L and C band coordinates are consistent with the {\it Gaia} positions to better than 6~mas, except for the extended component C2 in PG\,1426+015, which is about 40~mas away from the {\it Gaia} position.
We note that in PG\,0049+171 and PG\,1416$-$129, the L band position offsets, about 5.2--5.6~mas, are relatively large compared to the C band position offsets, about 0.3--0.4~mas.
In RL AGN, a relativistic jet may cause a ``core-shift'' effect, that is the centroid of the radio emission is shifted to the jet direction with frequency, due to Synchrotron opacity \citep[e.g.][]{Kovalev2017}.
However, these offsets are within the astrometric uncertainty, and we can not conclude whether they are caused by the core-shift effect.

The synthesized beam sizes and the deconvolved source sizes are also listed in Table~\ref{size}.
The sizes are measured in the full array maps, except for PG\,1116+215 in which the tapered map is used.
We note that in PG\,1244+026, the beam size in C band is larger than that in L band, which may be because the data in the C-band long baseline is mostly noise and is excluded in the fitting with DIFMAP.
If the deconvolved source size is smaller than half of the beam size, we consider the source as unresolved.

Table~\ref{flux} reports the total flux density $S_{\rm total}$, the core flux density $S_{\rm core}$, the background noise RMS, and the $uv$-range of the full-array and the tapered maps in the L and C bands.
For sources with only one component, we use the peak intensity, which is the unresolved flux density in a single beam, as the $S_{\rm core}$, and the $S_{\rm total}$ is the integrated flux density.
For sources with more than one component, we use the peak intensity of the core component as the $S_{\rm core}$, and the $S_{\rm total}$ is the sum of the integrated flux density of all components.
If the object is not detected, we use a 5$\sigma$ upper limit on $S_{\rm core}$.
In case of $S_{\rm total} < S_{\rm core}$, the source is unresolved, and we use $S_{\rm core}$ as an upper limit on $S_{\rm total}$.

The brightness temperature at 4.9~GHz is measured following
\begin{equation}
T_{\rm B} = 1.8 \times 10^{9} (1+z) \frac{S_\nu}{\nu^2 \theta_{\rm max} \theta_{\rm min}}
\end{equation}
\citep[e.g.][]{Ulvestad2005}, where $S_\nu$ is the total flux density in mJy, $\nu$ is the observing frequency in GHz, and $\theta_{\rm max}$ and $\theta_{\rm min}$ are the major and minor axes of the source size in mas.
If the emission is unresolved, we used the core flux density as $S_\nu$, and half of the beam size as $\theta_{\rm max}$ and/or $\theta_{\rm min}$, and thus the measured $T_{\rm B}$ is a lower limit.
The $T_{\rm B}$ of PG\,1116+215 is computed in the tapered map, and that of the two components in PG\,1426+015 is also given.
The values of $T_{\rm B}$ are also reported in Table~\ref{flux}.

The radio to X-ray luminosity ratio, $L_{\rm R}/L_{\rm X}$, is calculated using the EVN core flux density at 5~GHz and the X-ray flux at 0.2--12.0~keV from the {\it XMM-Newton} DR12 catalog \citep{Webb2020}.
If the object is not detected in our EVN observations, we set an upper limit on the ratio.
PG\,1012+008 is not detected in either radio or X-ray bands, so its ratio is unknown.
The radio to X-ray luminosity ratio is listed in Table~\ref{ratio}, which also includes the redshift, the physical scale, the BH mass from \citet{Davis2011}, the VLA A-configuration 5~GHz flux density from literature, and the X-ray 0.2--12~keV flux from the {\it XMM-Newton} DR12 catalog \citep{Webb2020}.
The values of the additional 13 RQQ detected with the VLBA can be found in \citet{Chen2023} and are also listed here.

The EVN spectral slope $\alpha_{\rm EVN}$ is measured based on the $S_{\rm total}$ in the tapered maps, which have comparable resolutions and cover emission on similar scales at both 1.7 and 4.9~GHz.
If the object is detected in only one band, the $S_{\rm core}$ is used to derive a limit on the slope.
We further measured the EVN core to total flux ratio $S_{\rm core}/S_{\rm total}$, and the ratio of the EVN total flux to the VLA A-configuration core flux \citep{Kellermann1989} $S_{\rm EVN}/S_{\rm VLA}$ both at 5~GHz.
For PG\,1426+015 which has two components, the $\alpha_{\rm EVN}$ is estimated using the $S_{\rm core}$ at 4.9~GHz and the 1.7~GHz upper limit of the two components, and the $S_{\rm core}/S_{\rm total}$ is the ratio of the core flux of the C1 component to the total flux of the C1+C2 components.
The $\alpha_{\rm EVN}$ and the $S_{\rm core}/S_{\rm total}$ of the individual components are also calculated.
The $S_{\rm EVN}/S_{\rm VLA}$ is the ratio of the EVN total flux of the two components to the VLA A-configuration core flux.
We note that the limits on the slopes need to be taken with caution because the non-detection images may have a higher noise level than our measurements.
The values of the additional 13 RQQ detected with the VLBA are calculated in the same way and can be found in \citet{Chen2023}.
Table~\ref{result} reports the EVN/VLBA spectral slope $\alpha_{\rm EVN/VLBA}$, the EVN/VLBA core to total flux ratio $S_{\rm core}/S_{\rm total}$, and the ratio of the EVN/VLBA total flux to the VLA A-configuration core flux $S_{\rm EVN/VLBA}/S_{\rm VLA}$ of the six RQQ detected in the EVN observations and the 13 RQQ detected in the VLBA observations.

We classified the objects with or without a BLR wind based on whether the C\,IV emission line profile shows strong excess blue wing emission compared to the H$\beta$ profile by an eye inspection (see Figures \ref{civ_evn} and \ref{civ_vlba}).
The origin of the radio emission is discussed in Section~\ref{classification}, and the classification is reported in Table~\ref{result}, which also includes the VLA spectral slope $\alpha_{\rm VLA}$ from literature, the Eddington ratio $\log L/L_{\rm Edd}$ from \citet{Davis2011}, and the C\,IV velocity shift in units of its FWHM from \citet{Baskin2005}.

\section{The origin of the radio emission} \label{classification}

The possible origin of the radio emission can be constrained based on the spectral slope and the compactness.
The compact flat-spectrum source is optically thick and possibly associated with the accretion disk corona.
The extended steep-spectrum emission is optically thin and likely produced by an AGN-driven wind or a low-power jet.
The brightness temperature in the EVN sample is about $\log T_{\rm B} = 5.7 - 6.8$ or higher, which is about one or two orders of magnitude higher than that is expected for SF and free-free emission \citep[$\ll 10^6$~K;][]{Njeri2024}.
This suggests that the pc-scale radio emission is probably AGN-driven.
We therefore focus on the origins of a wind, a jet, and the coronal emission as possible emission mechanisms.

The radio emission is associated with an outflow (including a wind and a jet) if the source shows extended radio morphology (i.e., more than one component), or meets two of the three below criteria: \\
-- steep $\alpha_{\rm EVN/VLBA}$ (i.e., $< -0.5$), \\
-- steep $\alpha_{\rm VLA}$ (i.e., $< -0.5$), \\
-- low $S_{\rm core}/S_{\rm total}$ (i.e., $< 0.5$). \\
In contrast, the radio source is considered to be compact, and therefore likely to have a coronal origin, if it meets two of the three below criteria: \\
-- flat $\alpha_{\rm EVN/VLBA}$ (i.e., $> -0.5$), \\
-- flat $\alpha_{\rm VLA}$ (i.e., $> -0.5$), \\
-- high $S_{\rm core}/S_{\rm total}$ (i.e., $> 0.5$). \\

In three of the six objects detected with the EVN, PG\,1116+215, PG\,1244+026, and PG\,1404+226, the radio emission may be associated with an AGN wind.
In the other three objects, the radio emission may be of coronal origin in PG\,0049+171 and PG\,1416$-$129, and a jet or a collimated outflow may be present in PG\,1426+015.
A detailed discussion of individual objects can be seen in Subsections 5.1--5.6.

The above criteria are used to separate the corona from the outflow, and can not tell a wind and a jet apart given their similarity in terms of morphology and spectral slope.
Additional information is needed to distinguish between a wind and a jet, for instance, the jet may have a higher $T_{\rm B}$ than the wind due to the Doppler effect.
An earlier VLBA study \citep{Chen2023} suggests that the $T_{\rm B}$ of the extended wind emission is generally lower, about $10^6 - 10^7$~K, than that of the corona and the jet emission, which spreads around $10^6 - 10^9$~K.
However, we note that the $T_{\rm B}$ based on the EVN measurements is systematically lower than that based on the VLBA measurements, which is probably due to the different resolutions.
We thus can not compare the values measured in different arrays, but we can compare them measured in the same array.
Indeed, the $T_{\rm B}$ of the radio wind objects ($10^{5.7}-10^{6.4}$~K) is on average lower than that of the corona or jet sources ($10^{6.5}-10^{6.8}$~K).
The value of $S_{\rm EVN/VLBA}/S_{\rm VLA}$ has to be taken with caution, because the object may vary among the non-simultaneous observations, especially for the compact sources, and thus it is not considered in the outflow and corona criteria.

\subsection{PG\,1116+215: Wind}

PG\,1116+215 was detected only after tapering, thus the radio emission in this object is mostly extended ($\gtrsim 10$~pc), with $S_{\rm core}/S_{\rm total} = 0.4 - 0.5$ at both 5 and 1.7~GHz.
Significant extended emission is dominated on larger scales, since $S_{\rm EVN}/S_{\rm VLA} = 0.2$ at 5~GHz.
The EVN and the VLA slopes are both steep with $\alpha_{\rm EVN} = -1.18$ at 1.7--4.9~GHz and $\alpha_{\rm VLA} = -0.59$ at 5--8.5~GHz \citep{Laor2019}.
The object meets all three of the outflow criteria, suggesting optically thin emission likely from a wind.

\subsection{PG\,1244+026: Wind}

PG\,1244+026 is unresolved at 5~GHz ($< 3.5$~pc), with $S_{\rm core}/S_{\rm total} = 1$, but extended emission is present at 1.7~GHz, where $S_{\rm core}/S_{\rm total} = 0.6$, although this frequency happens to have a higher resolution ($\sim 2.5$~pc).
Significant 5~GHz extended emission is present on larger scales ($\sim 300$~pc), as $S_{\rm EVN}/S_{\rm VLA} = 0.4$.
The EVN slope, $\alpha_{\rm EVN} = -0.52$, is close to the flat versus steep dividing line, which may suggest comparable contributions at 1.7--5~GHz from a compact flat source which dominates above 5~GHz and a steep extended source which dominates below 1.7~GHz.
The VLA slope of larger scale emission is indeed steep with $\alpha_{\rm VLA} = -0.87$ at 1.4--5~GHz \citep{Jarvela2022}.
The object meets two of the three outflow criteria, indicating the presence of a wind.

\subsection{PG\,1404+226: Wind}

PG\,1404+226 is dominated by extended emission at 5~GHz ($> 3.5$~pc), with $S_{\rm core}/S_{\rm total} = 0.4$, and remains extended at 5~GHz up to the VLA scale ($\sim 600$~pc), where $S_{\rm EVN}/S_{\rm VLA} = 0.5$.
The VLA slope is steep with $\alpha_{\rm VLA} = -0.63$ at 5--8.5~GHz \citep{Laor2019}.
The object meets two of the three outflow criteria, suggesting extended radio emission from a wind.
Though the source is extended, the non-detection at 1.7~GHz gives a flat EVN slope, $\alpha_{\rm EVN} > 0.09$, which may be due to free-free absorption (see Section~\ref{FFA}).

\subsection{PG\,0049+171: Corona}

PG\,0049+171 is mostly unresolved ($< 2.6$~pc) at both 5 and 1.7~GHz, with $S_{\rm core}/S_{\rm total} = 0.8$ in both bands.
The EVN and the VLA slopes are both flat, $\alpha_{\rm EVN} = 0.13$ at 1.7--4.9~GHz and $\alpha_{\rm VLA} = -0.3$ at 5--8.5~GHz \citep{Laor2019}, suggesting highly compact ($< 0.1$~pc) optically thick emission.
The object meets all three of the corona criteria, which indicates that the radio emission is likely originated from the corona.
The low EVN/VLA flux ratio, $S_{\rm EVN}/S_{\rm VLA} = 0.4$, may result from a factor of $\sim 2$ flux variability, or extended emission on larger scales.

\subsection{PG\,1416$-$129: Corona}

PG\,1416$-$129 shows mostly compact ($< 6.2$~pc) emission, as $S_{\rm core}/S_{\rm total} = 1$ at 5~GHz and $S_{\rm core}/S_{\rm total} = 0.8$ at 1.7~GHz.
The EVN slope $\alpha_{\rm EVN} = -0.59$ at 1.7--4.9~GHz is close to the flat versus steep dividing line, which may again result from a transition from steep extended emission at lower frequencies to flat compact emission at higher frequencies, as indicated by the VLA slope $\alpha_{\rm VLA} = 0.11$ at 5--8.5~GHz \citep{Barvainis1996}.
The object meets two of the three corona criteria, which suggests that the compact emission at 5~GHz is likely dominated by the coronal emission.
The $S_{\rm EVN}/S_{\rm VLA} = 0.5$ may suggest the presence of extended emission on larger scales, or a factor of $\sim 2$ flux variability.

\subsection{PG\,1426+015: Jet}

PG\,1426+015 has two components in the EVN observations at 4.9~GHz, which suggests the object launches an outflow (a wind or a jet).
The C1 component is unresolved with $S_{\rm core}/S_{\rm total} = 1$ and the C2 component is extended with $S_{\rm core}/S_{\rm total} = 0.5$.
Significant extended emission is likely present on larger scales, as $S_{\rm EVN}/S_{\rm VLA} = 0.5$.
The non-detection in the L band results in flat EVN slopes in both components.
The total EVN slope of both components are flat with $\alpha_{\rm EVN} > -0.07$ at 1.7--4.9~GHz, which is consistent with the VLA slope of $\alpha_{\rm VLA} = -0.18$ at 5--8.5~GHz \citep{Laor2019} within the uncertainty.
The $T_{\rm B}$ of C2 is comparable to that of C1, and is relatively high compared to the other three wind sources.
This may favor that the radio emission is associated with a jet or a compact outflow, with a projected size of about 68.3~pc away from the AGN center, on the NLR scale.
The C1 component is probably associated with the corona or the jet base with a flat slope ($\alpha_{\rm EVN} > -0.40$).
The flat slope of the C2 component ($\alpha_{\rm EVN} > 0.17$) may be also due to free-free absorption in the outflow gas (see Section~\ref{FFA}). 
However, the wind interpretation can not be completely ruled out.
Further study, such as proper motion, will help to clarify its radio origin.

\subsection{EVN non-detections}

Four of the ten objects in our sample were not detected in the EVN observations.
The $5\sigma$ upper limits on their 5~GHz flux are 0.10--0.14~mJy (Table~\ref{flux}), which are lower than the range of detected fluxes of 0.18--0.39~mJy.
Are these four objects unusually radio weak, or are they typical RQQ which just happen to be more distant or less luminous?
Their redshift distribution $0.081 \le z \le 0.459$ is clearly higher than that of the detected objects $0.048 \le z \le 0.176$.
Their $\log L_{\rm R}/L_{\rm X}$ upper limits also fall within the range spanned by the detected objects (Table~\ref{ratio}).
Thus the undetected objects are not necessarily radio weak and may have a typical radio to X-ray flux ratio for RQQ.

\section{The radio and the BLR winds}

We present our search for the correlations between the radio evidence for the pc-scale winds and the C\,IV evidence for the BLR winds on $\sim$ 0.01--0.1~pc scales.
To improve the statistics, we further include additional 13 RQ PG quasars, which are detected with the VLBA at 1.6--4.9~GHz \citep{Alhosani2022,Chen2023} and have an observed C\,IV emission line profile \citep{Baskin2005}.
The classification of the radio emission origin of these 13 objects is discussed in \citet{Alhosani2022} and \citet{Chen2023}.
Combining the EVN and VLBA observations, we have a sample of 19 objects totally.

The objects are divided into four groups based on whether the object has radio and/or BLR winds: \\
-- Radio + BLR winds (five objects): the radio emission is associated with a wind, and the object also has a BLR wind; \\
-- Radio wind only (four objects): the radio emission is associated with a wind, but the object does not have a BLR wind; \\
-- BLR wind only (one object): the object has a BLR wind, but the radio emission is associated with the corona or a jet; \\
-- No wind (nine objects): the object does not have a BLR wind, and the radio emission is associated with the corona or a jet. \\
A detailed grouping is listed in Table~\ref{result}.

\begin{figure}[ht!]
\centering
\includegraphics[width=\columnwidth, trim={0cm, 0cm, 1cm, 1cm}, clip]{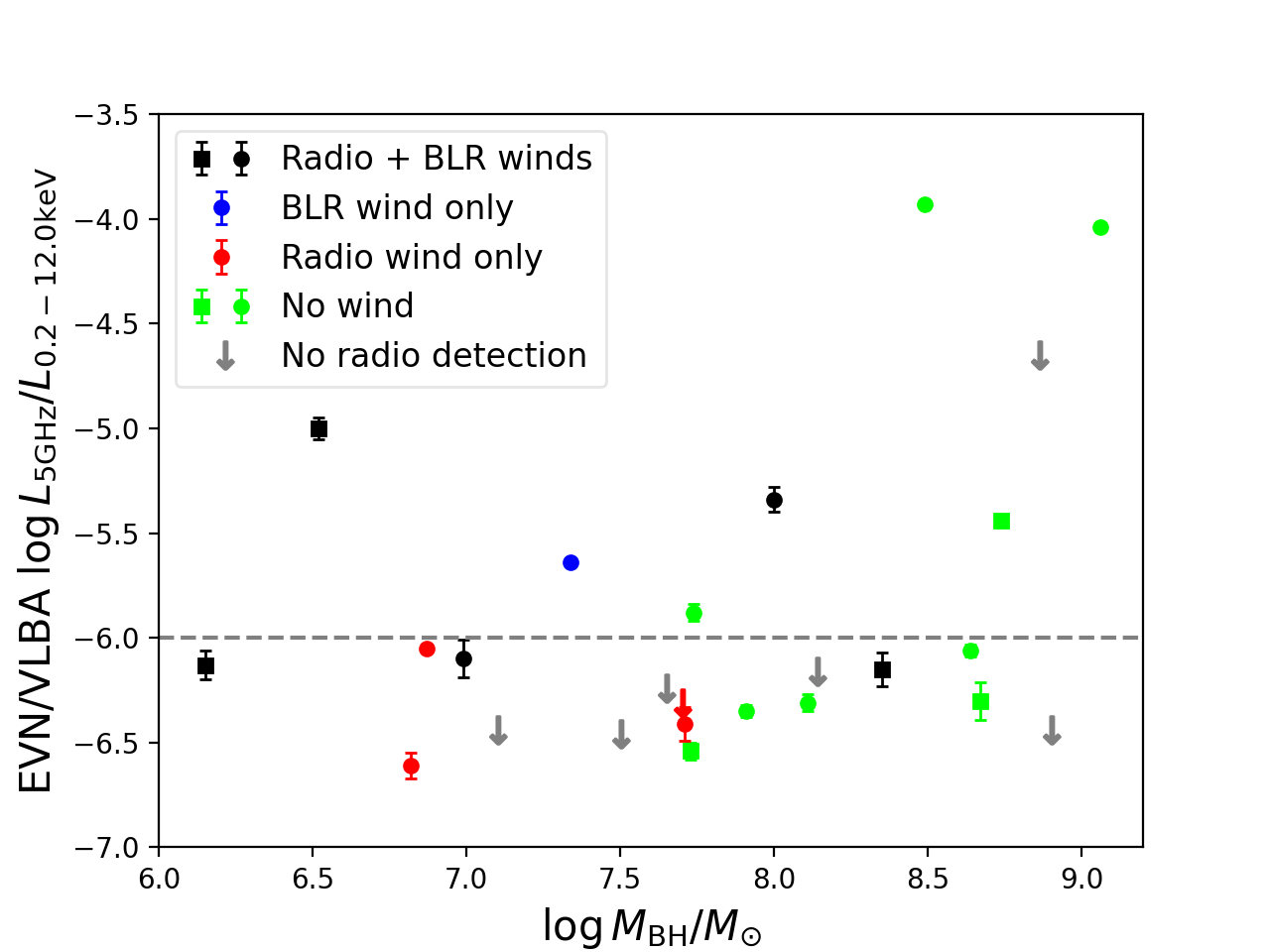}
\caption{The radio 5~GHz to X-ray 0.2--12~keV luminosity ratio as a function of the BH mass.
The objects are divided into 4 groups based on showing both radio and BLR winds (black), either BLR wind (blue) or radio wind (red) only, or neither BLR nor radio winds (green, including corona and jet in radio).
The classification can be seen in Table~\ref{result}.
The squares and circles represent the objects observed with EVN and VLBA respectively.}
\label{Lrx+Mbh}
\end{figure}

\begin{figure}[ht!]
\centering
\includegraphics[width=\columnwidth, trim={4cm, 0cm, 6cm, 2cm}, clip]{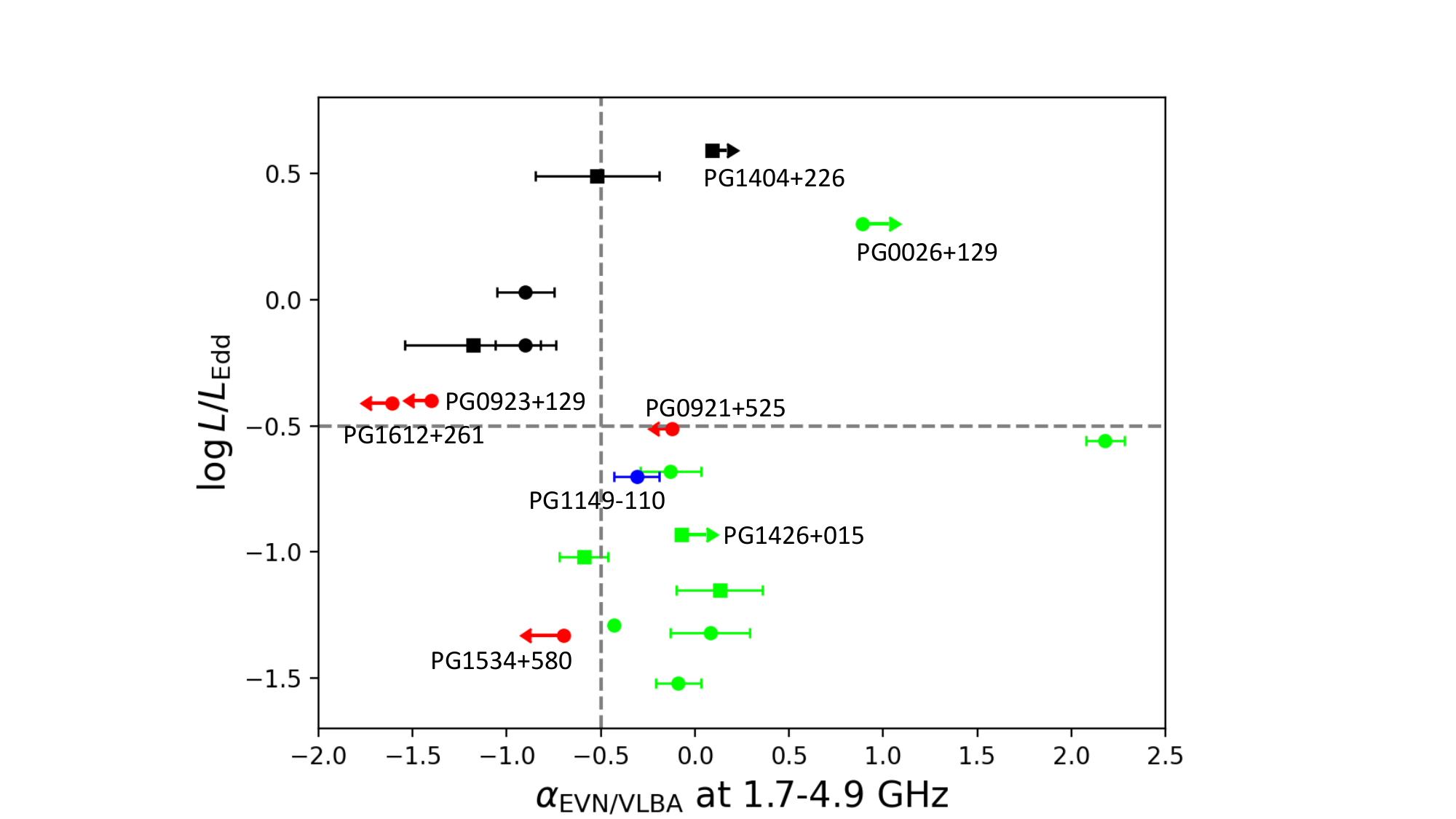}
\caption{The Eddington ratio as a function of the radio slope at 1.7--4.9~GHz.
The symbols are the same as Figure~\ref{Lrx+Mbh}. The name of the objects discussed in Section~\ref{discuss} is labeled.
PG\,1404+226 is likely affected by free-free absorption, and its intrinsic slope could be steep (see Section~\ref{FFA}).
The five objects with both radio and BLR winds (black) reside at $\log L/L_{\rm Edd} > -0.5$, while eight of the nine objects without a wind (green) reside at $\log L/L_{\rm Edd} < -0.5$.}
\label{Edd+slope}
\end{figure}

\begin{figure}[ht!]
\centering
\includegraphics[width=\columnwidth, trim={4cm, 0cm, 6cm, 2cm}, clip]{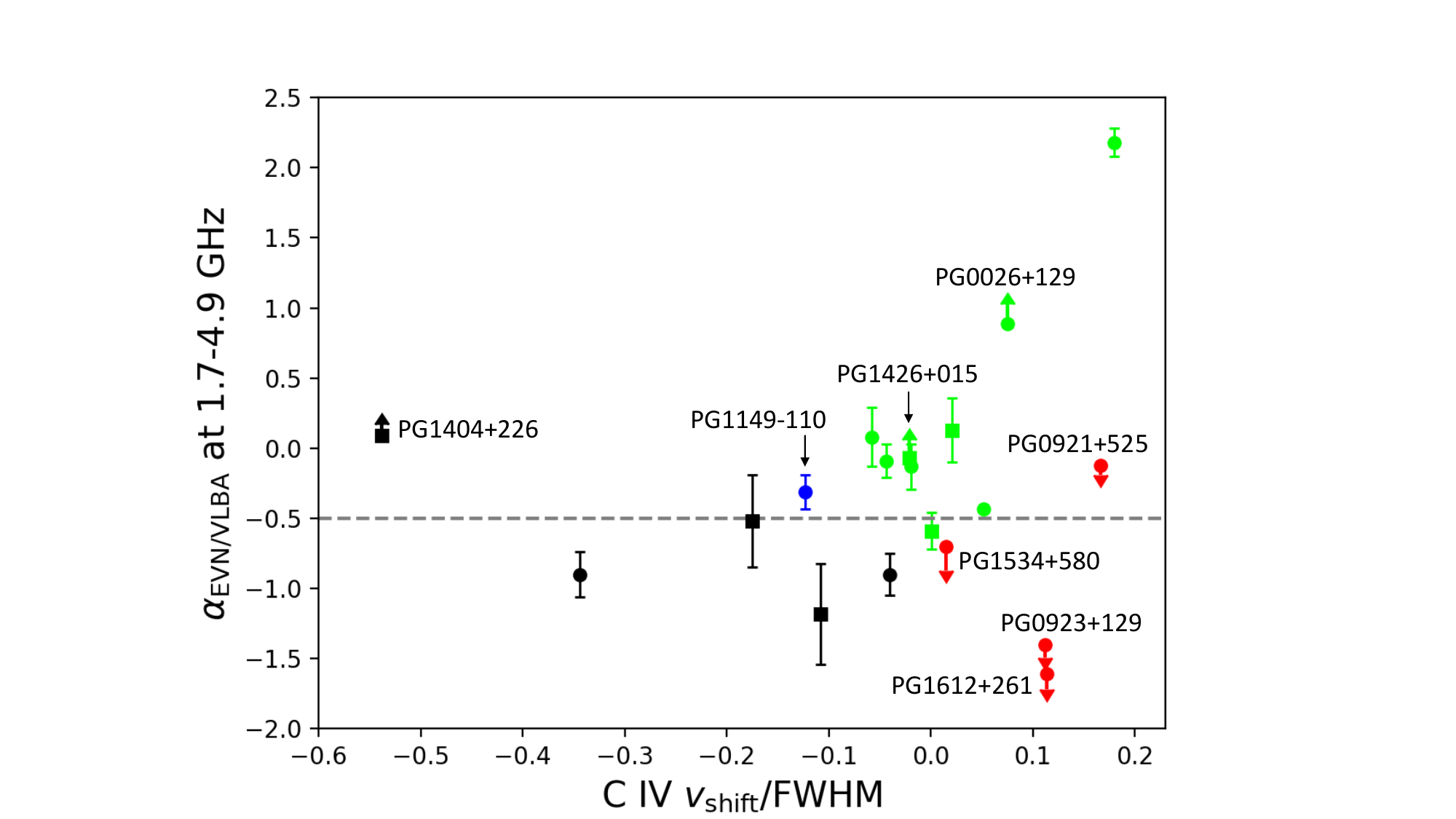}
\caption{The radio slope at 1.7--4.9~GHz as a function of the C\,IV velocity shift divided by its FWHM.
The symbols are the same as Figure~\ref{Lrx+Mbh}. The name of the objects discussed in Section~\ref{discuss} is labeled.
PG\,1404+226 is likely affected by free-free absorption, and its intrinsic slope could be steep (see Section~\ref{FFA}).
The five objects with both radio and BLR winds (black) show a strong C\,IV blue excess, while the nine objects without a wind (green) do not show.}
\label{slope+shift}
\end{figure}

\begin{figure}[ht!]
\centering
\includegraphics[width=\columnwidth, trim={4cm, 0cm, 6cm, 2cm}, clip]{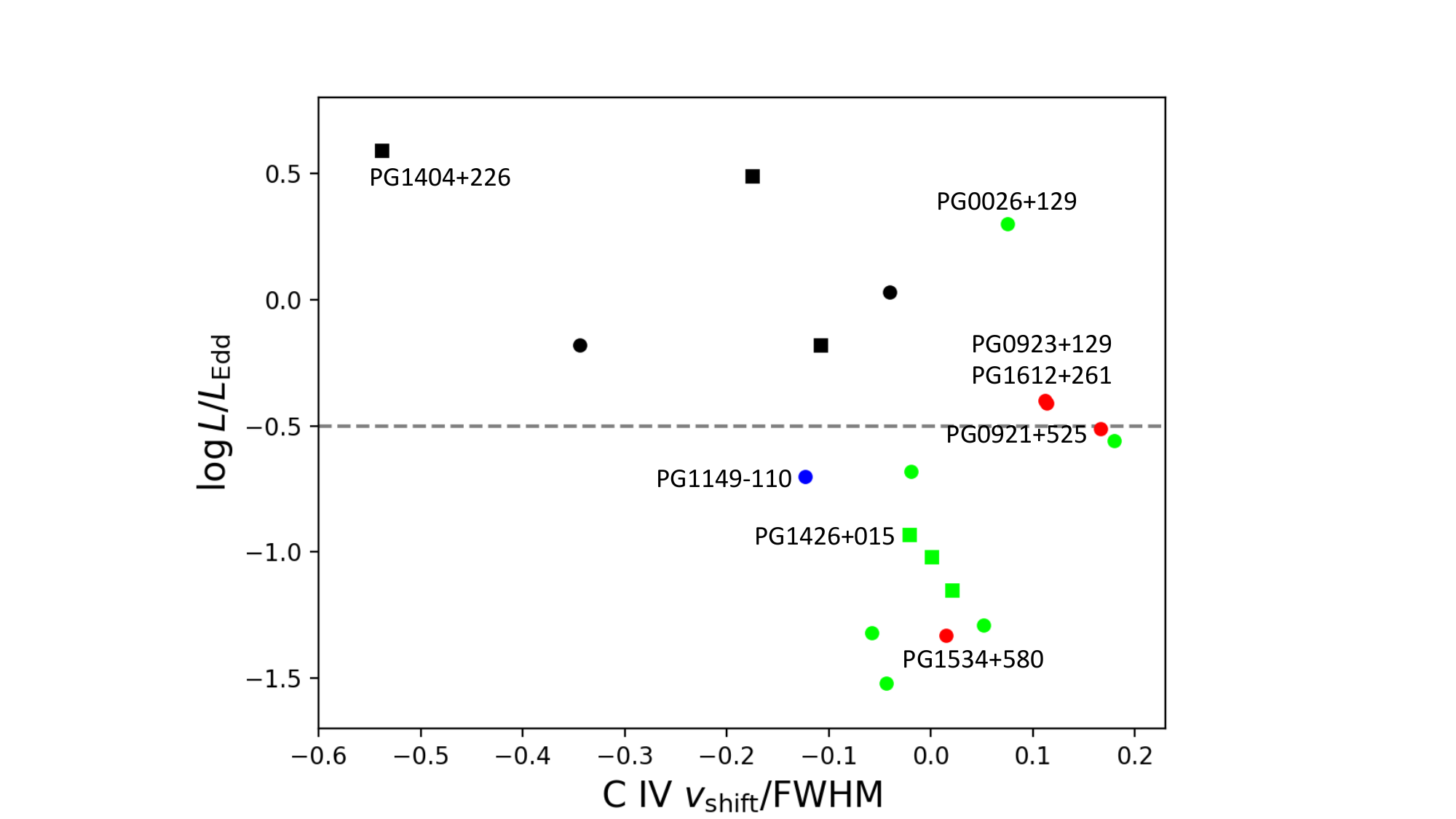}
\caption{The Eddington ratio as a function of the C\,IV velocity shift divided by its FWHM.
The symbols are the same as Figure~\ref{Lrx+Mbh}. The name of the objects discussed in Section~\ref{discuss} is labeled.}
\label{Edd+shift}
\end{figure}

Figure~\ref{Lrx+Mbh} presents the ratio of the EVN or VLBA core luminosity at 5~GHz to the X-ray luminosity at 0.2--12~keV.
Most of the objects cluster around $\log L_{\rm R}/L_{\rm X} \sim -6$, which is typical for RQ AGN without a powerful relativistic jet \citep{Fischer2021,Chen2023}.
The radio + BLR wind objects are spread over $\log M_{\rm BH}/M_{\odot} = 6.2 - 8.4$, while the no wind objects reside only at $\log M_{\rm BH}/M_{\odot} > 7.7$.
This may be an $L/L_{\rm Edd}$ effect, where high $L/L_{\rm Edd}$ objects which can launch a radiation pressure driven wind, are found at all BH masses, while low $L/L_{\rm Edd}$ objects which do not drive a wind, are more common at high BH masses.

Figure~\ref{Edd+slope} shows the $L/L_{\rm Edd}$ versus $\alpha_{\rm EVN/VLBA}$ distribution.
In the five radio + BLR wind objects, 4/5 have high Eddington ratios ($\log L/L_{\rm Edd} > -0.5$) and steep spectral slopes ($\alpha_{\rm EVN/VLBA} < -0.5$).
In the nine objects with neither radio nor BLR winds, 8/9 (within the uncertainty) have low Eddington ratios ($\log L/L_{\rm Edd} < -0.5$) and flat spectral slopes ($\alpha_{\rm EVN/VLBA} > -0.5$).
Thus about 86\% of the objects are consistent with a general interpretation that the winds are characterized by a high $L/L_{\rm Edd}$, a steep $\alpha_{\rm EVN/VLBA}$, and strong C\,IV excess blue wing, and vice versa.

The Spearman correlation suggests a weak trend between $L/L_{\rm Edd}$ and $\alpha_{\rm EVN/VLBA}$ with $r = -0.57$ and $p = 0.05$ when excluding objects with a slope limit.
The Kolmogorov–Smirnov (KS) tests suggest that the $L/L_{\rm Edd}$ in the objects with or without BLR wind, and in the objects with or without radio wind, are both drawn from different populations at a confidence level of 98.9\% ($p$ = 0.011) for BLR wind and of 99.7\% ($p$ = 0.003) for radio wind.
The KS tests suggest that the $\alpha_{\rm EVN/VLBA}$ can distinguish the objects with or without pc-scale radio wind at a confidence level of 98\% ($p$ = 0.020), but it can not tell the objects with or without BLR wind apart.

Figures \ref{slope+shift} and \ref{Edd+shift} show the distributions of $\alpha_{\rm EVN/VLBA}$ and $L/L_{\rm Edd}$, respectively, against C\,IV velocity shift to its FWHM which indicates the strength of the BLR wind.
The C\,IV emission line tends to have a relatively large blueshift in the five radio + BLR wind objects, where all have high $L/L_{\rm Edd}$ and 4/5 have steep $\alpha_{\rm EVN/VLBA}$.
The nine no wind objects tend to have no blueshift or even show a redshift, where 8/9 have low $L/L_{\rm Edd}$ and all (within the uncertainty) have flat $\alpha_{\rm EVN/VLBA}$.

The Spearman correlation suggests the trend between $\alpha_{\rm EVN/VLBA}$ and C\,IV shift is even weaker with $r = 0.50$ and $p = 0.09$ when excluding the slope limits, and no trend between $L/L_{\rm Edd}$ and C\,IV shift.
The KS tests suggest that the C\,IV shift can distinguish the objects with or without BLR wind at a confidence level of 99.8\% ($p$ = 0.002), but it can not tell the objects with or without radio wind apart.

In total, about 74\% (14/19) of the objects show either both radio and BLR winds or neither, in which, about 86\% (12/14 within the uncertainty) of the sources agree with the interpretation that the wind objects are characterized by high $L/L_{\rm Edd}$ and steep $\alpha_{\rm EVN/VLBA}$, and the no wind objects are characterized by low $L/L_{\rm Edd}$ and flat $\alpha_{\rm EVN/VLBA}$.
The Eddington ratio could be a good indicator for the presence of an AGN wind, which is driven by the radiation pressure, as it can distinguish the groups with or without radio and BLR winds better than the other parameters based on the KS tests.
High $L/L_{\rm Edd}$ objects are likely to launch a strong wind, which produces a blueshifted component in the C\,IV emission line from the BLR, as well as the optically thin radio emission by an interaction with the ambient medium on the NLR scales.
In contrast, it appears that neither a BLR nor a NLR wind is launched in low $L/L_{\rm Edd}$ objects, and thus no blueshifted component is seen in the C\,IV emission line, and the optically thick radio emission is likely originated from the corona.

\section{Discussion} \label{discuss}

The relation of the wind phenomenon with the $L/L_{\rm Edd}$ suggests that the wind is likely radiation pressure driven.
High $L/L_{\rm Edd}$ objects are capable of producing the winds, both in the BLR as indicated by the excess blue wing in the C\,IV emission line, and in the NLR as indicated by the pc-scale extended steep-spectrum radio emission.
The winds are not detected in low $L/L_{\rm Edd}$ objects in either the BLR or the NLR, as demonstrated by the symmetric C\,IV profile and the compact flat-spectrum radio emission.

However, there are exceptions.
We here discuss various physical effects which can affect the relations between the radio and the BLR wind indicators presented earlier.

\subsection{Free-free absorption} \label{FFA}

The radio emission can be absorbed by the AGN photoionized gas via free-free absorption, which is set by the AGN ionizing luminosity and by the distance of the absorber from the AGN \citep{Baskin2021}.
In this case, the spectral slope will become flatter, and possibly inverted, at lower frequencies where the free-free absorption dominates.

PG\,1404+226 has a high $L/L_{\rm Edd}$ and the signature of both radio and BLR winds.
However, the non-detection in our EVN observations at 1.7~GHz implies a flat $\alpha_{\rm EVN}$ ($> 0.09 \pm 0.08$), in contrast with most of the other high $L/L_{\rm Edd}$ objects where $\alpha < -0.5$ (Figure~\ref{Edd+slope}).
If the flat $\alpha_{\rm EVN}$ is produced by an optically thick Synchrotron source, the emission must be compact \citep[$< 0.1$~pc;][]{Laor2008} and mostly unresolved.
However, a large fraction of the 5~GHz emission is resolved ($S_{\rm core}/S_{\rm total} = 0.4$), i.e., on a scale of larger than $1.87 \times 5.97$~mas$^2$ or $3.5 \times 11.3$~pc$^2$.
This extended emission is inevitably steep, and should have been detected at 1.7~GHz at about 10$\sigma$ or higher, which is in contrast with the observation.
We suggest that the observed flat spectrum may be caused by free-free absorption.

PG\,1404+226 shows a narrow C\,IV absorption line of intermediate-strength \citep[an absorption equivalent width of 1.5~{\AA};][]{Laor2002}, which indicates a low-velocity wind ($\sim 2000$~km~s$^{-1}$) along our line of sight.
The radio emission can be free-free absorbed by the outflowing gas when passing through it.
Does the free-free absorption from such a wind produce significant spectral flattening below 5~GHz?

The free-free absorption frequency of photoionized gas allows to estimate the distance of the absorber.
If the observed radio emission is indeed optically thick at 1.7~GHz and optically thin at 4.9~GHz, then the optical depth $\tau_{\rm ff} > 1$ at 1.7~GHz and likely $\tau_{\rm ff} < 1$ at 4.9~GHz.
For simplicity, let us assume $\tau_{\rm ff} = 1$ at 3~GHz.
The location of the absorber, $r$, can be estimated from the following relation \citep[eq.21 in][]{Baskin2021}
\begin{equation}
\nu_{\rm thick} = 6.03 \times 10^{11} (\frac{r}{r_{\rm dust}})^{-0.95} \, \rm{Hz}
\end{equation}
for dusty gas.
Here $\nu_{\rm thick}$ is the frequency at which $\tau_{\rm ff} = 1$ and assumed to be 3~GHz.
The dust sublimation radius, $r_{\rm dust}$, is defined as \citep[eq.11 in][]{Baskin2021}
\begin{equation}
r_{\rm dust} = 0.2 L_{46}^{0.5} \, \rm{pc}
\end{equation}
where $L_{46}$ is the bolometric luminosity in units of $10^{46}$\,erg\,s$^{-1}$, and $\log L_{\rm bol} = 45.21$ for this object \citep{Davis2011}.
The derived distance of the absorbing medium from the central source is then 15.7~pc, which is comparable to the projected source size of 11.4 $\times$ 4.9~pc$^2$.
Thus the UV wind, if its lateral dimensions are comparable to its distance from the source, could be large enough to cover the radio source, and its free-free absorption can produce the observed inverted $\alpha_{\rm EVN}$.

Free-free absorption may be also affecting PG\,1426+015, which shows two emission components.
The component C1 is unresolved and shows flat-spectrum emission, which coincides well with the {\it Gaia} position.
It is therefore most likely the core emission, probably the corona or the jet base.
The component C2 resides 68~pc away in projection from the core.
Its slope is flat, but its emission is significantly spatially resolved ($S_{\rm core}/S_{\rm total} = 0.5$), in contrast to an optically thick Synchrotron source which is compact and remains unresolved.
C2 is similar to PG\,1404+226, by also being very flat ($\alpha_{\rm EVN} > 0.17 \pm 0.14$), yet significantly extended (free-free emission is excluded by the high $T_B$ of both sources).
The likely interpretation in this case is therefore also a free-free absorption screen which resides in front of C2.

Is a photoionized absorbing screen in PG\,1426+015 expected to be optically thick to free-free absorption at the range of 1.5--5~GHz?
The bolometric luminosity is $\log L_{\rm bol}$ = 45.84 for this object \citep{Davis2011}, and the absorber distance is $\ge 68$~pc (the projected separation between C1 and C2).
Following the derivation earlier, it indeed gives $\nu_{\rm thick} = 2.0$~GHz, as required.
Specifically, since $\tau_{\rm ff} \propto \nu^{-2}$ in the radio regime \citep[e.g.][]{Rybicki1986}, we get $\tau_{\rm ff}$(1.5GHz) = 1.78 and $\tau_{\rm ff}$(5GHz) = 0.16, and the absorption corrected slope indeed becomes steep with $\alpha = -1.18$.

In contrast with PG\,1404+226 which shows associated C\,IV absorption, in PG\,1426+015 no C\,IV absorption is detected \citep{Laor2002}.
This is because in PG\,1404+226 the free-free absorption screen appears to cover the core, while in PG\,1426+015 the core, where the C\,IV line is emitted, is not absorbed in the radio.
The free-free absorption screen in PG\,1426+015 resides $\ge 68$~pc from the nucleus, and is likely associated with photoionized gas in the NLR.

\subsection{The origin of the extended radio emission}

The extended radio emission may originate in a jet lobe or a wind, which both produce extended steep-spectrum radio emission.
In principle, a jet is expected to be highly collimated compared to a wide-angle wind.
However, in practice, it is generally difficult to separate a jet from a wind when the emission is not well resolved in the radio observations.

Only PG\,1426+015 shows a separated extended component (C2) in addition to the core component (C1).
The projected distance of 68.3~pc between C1 and C2 suggests that C2 resides on the NLR scale.
The angular scale of C2 is 4.66~mas at the distance of 40.4~mas from C1, which corresponds to a half opening angle of only 3.3$^{\circ}$.
Since the NLR gas likely extends over a large solid angle, the small opening angle of C2 suggests a well-collimated outflow, i.e., a jet rather than a wind.
In addition, the $T_{\rm B}$ of C2 is comparable to that of C1, and is relatively high compared to the other three wind sources in the EVN observations, which may also support the jet interpretation.

We note in passing that the HST image of PG\,1426+015 shows an additional source in the inner part of the host galaxy with a separation of $\sim$ 2~arcsec from the central nucleus in the same direction of C2 \citep{Bentz2009a}.
The C2 component may thus be related to the interaction between the two objects.
However, the corresponding distance between the two objects is about 50 times larger than the C1 and C2 separation ($\sim 40$~mas), which cannot be resolved with the HST.

\subsection{The association of the radio and BLR winds}

The five objects with both radio and BLR winds (PG\,0050+124, PG\,0157+001, PG\,1116+215, PG\,1244+026, and PG\,1404+226) have a very high $L/L_{\rm Edd}$ ($\ge$ 0.66), which suggests that the winds are radiation pressure driven.
There are four objects, PG\,0921+525, PG\,0923+129, PG\,1534+580, and PG\,1612+261, with a radio outflow but without a BLR wind, and one object, PG\,1149$-$110, with a BLR wind but without a radio outflow.
They have a lower $L/L_{\rm Edd}$ than the objects with both radio and BLR winds.

PG\,0921+525 shows one-sided radio extended emission both on the VLBA scale \citep{Chen2023} and on the VLA A-configuration scale \citep{Kukula1998}.
Extended components are seen in PG\,0923+129 (two-sided) and PG\,1534+580 (one-sided) on the VLBA scale \citep{Chen2023}, but are not detected on the VLA A-configuration scale \citep{Leipski2006,Berton2018}.
PG\,1149$-$110 and PG\,1612+261 show two-sided radio extended emission on the VLA A-configuration scale \citep{Leipski2006}, but an extended component is not detected on the VLBA scale \citep{Alhosani2022,Wang2023a}.

Interestingly, four of the five objects with either a radio or a BLR wind (PG\,0921+525, PG\,0923+129, PG\,1149$-$110, and PG\,1612+261) are situated at a narrow range of $L/L_{\rm Edd}$ (0.20--0.40), just below the $L/L_{\rm Edd}$ in the radio + BLR wind objects.
The intermediate $L/L_{\rm Edd}$ may be sufficient to drive a mild wind.
In PG\,1534+580, the $L/L_{\rm Edd}$ (0.05) is significantly low, and likely not sufficient for a radiation pressure driven wind.
The extended radio emission may be associated with a weak jet.

\subsection{Inclination effects}

Of the nine objects without a wind indicator in either the radio or the BLR, eight reside at $L/L_{\rm Edd} \le 0.28$, which supports the scenario that the radiation pressure driven wind occurs only in the highest $L/L_{\rm Edd}$ objects.
The significant outlier in this group is PG\,0026+129, which shows neither a radio nor a BLR wind, despite of its very high $L/L_{\rm Edd}$ (2.0).
Can the apparently high $L/L_{\rm Edd}$ be a spurious inclination effect?
A pole-on orientation of a disk-like BLR will underestimate the H$\beta$ line width, and thus the BH mass, which results in an overestimate of $L/L_{\rm Edd}$.

PG\,0026+129 has an inverted $\alpha_{\rm VLBA}$ ($> 0.89 \pm 0.14$), compact emission ($S_{\rm core}/S_{\rm total} = 0.7$), a flat slope also on a lower resolution ($\alpha_{\rm VLA} = -0.31$), and shows variability ($S_{\rm VLBA}/S_{\rm VLA} = 1.6$), which are indicative of the highly compact radio emission.
Similarly, no C\,IV blueshifted absorption or emission lines suggests the absence of a BLR wind.
This object has a narrow H$\beta$ line with FWHM = 1860~km\,s$^{-1}$ and a relatively low-mass BH with $\log M_{\rm BH}/M_{\odot} = 7.7$ \citep{Davis2011}.
The HST image indeed shows that it is close to a face-on view \citep{Bentz2009a}.
Thus the $L/L_{\rm Edd}$ may potentially be overestimated due to a close to face-on view effect.

In addition, inclination can also affect the observed C\,IV line wind indicator.
The BLR wind is likely driven up out of the accretion disk by radiation pressure, and it may be close to the plane of the disk with an opening angle of a few degrees from the disk plane \citep{Murray1995,Proga2000}.
This wind will produce blueshifted absorption lines when viewed through the wind, and also blueshifted emission lines which will be enhanced in a close to edge-on view.
Thus a face-on view may be less likely to see the C\,IV line absorption and emission wind indicators, and vice versa.
In contrast, the slope and the compactness of the radio emission are less likely to be inclination-dependent, as there is no evidence for highly relativistic outflow in RQQ.

A close to face-on view of a disk wind is therefore another possible explanation for the presence of a radio outflow but the absence of a BLR wind in the four intermediate objects (PG\,0921+525, PG\,0923+129, PG\,1534+580, and PG\,1612+261).
The HST images of PG\,0921+525, PG\,0923+129, and PG\,1534+580 show a close to face-on host galaxy, suggesting that they are indeed viewed at a small inclination \citep{Bentz2009a,Bentz2013}.
In contrast, the host galaxy of PG\,1149$-$110 is highly inclined \citep{Zhao2021}.
A close to edge-on view of the disk wind may facilitate the detection of a BLR wind.

If the inclination interpretation is correct, then spectropolarimetry of the likely close to face-on objects should reveal low continuum polarization percentages $\%P$, while the BLR line profiles in polarized light, which reflects a close to edge-on view, may reveal the blueshifted wind component \citep{Capetti2021}.
Indeed, the white light polarimetry survey \citep{Berriman1990} find a low $\%P$ polarization in PG\,0026+129 ($0.27 \pm 0.17$), which is an outlier with a very high $L/L_{\rm Edd}$ (2.0) but without a wind, supporting the face-on view bias.
A low $\%P$ also characterizes the three intermediate $L/L_{\rm Edd}$ (0.3--0.4) objects, PG\,0921+525 ($0.17 \pm 0.08$), PG\,0923+129 ($0.12 \pm 0.17$), and PG\,1612+261 ($0.07 \pm 0.13$), which again supports the close to face-on bias.

A relatively high $\%P$ polarization is found in PG\,1534+580 ($0.79 \pm 0.14$), which argues against the inclination bias, and the BLR wind is likely absent as suggested by its very low $L/L_{\rm Edd}$ (0.05).
The five very high $L/L_{\rm Edd}$ ($\ge$ 0.66) objects with both radio and BLR winds, PG\,0050+124 ($0.61 \pm 0.08$), PG\,0157+001 ($1.37 \pm 0.40$), PG\,1116+215 ($0.23 \pm 0.11$), PG\,1244+026 ($0.48 \pm 0.25$), and PG\,1404+226 ($0.37 \pm 0.35$), and the one intermediate $L/L_{\rm Edd}$ (0.2) object, PG\,1149$-$110 ($0.23 \pm 0.11$), tend to have a relatively high $\%P$ and may be not biased by inclination, though the S/N is too low for a definite conclusion.

\subsection{Variability}

Throughout this study, we make use of public data from literature spanning a few decades, including the UV and optical spectroscopy, optical spectropolarimetry, past radio observations with the VLA A configuration, and the {\it XMM-Newton} catalog.
We note that variability may affect the results at a certain level, due to the variations of emission line profiles and the non-simultaneous radio observations.
However, variability, in general, will cause the data points to move randomly, which tends to weaken or destroy the trends.
Therefore, if the correlations still hold despite of variability, the correlations would be even stronger without variability.
Future new observations would be beneficial to confirm or disprove these results.

\section{Summary}

In this work, we look for the pc-scale radio emission associated with a wind, and explore its relation with the BLR wind indicated by the excess blue wing in the C\,IV emission line in 19 RQ PG quasars.
In the sample, six objects are from our new EVN observations (10 observed) at 1.7 and 4.9~GHz, and 13 objects are from our earlier VLBA studies (18 observed) at the same frequencies \citep{Alhosani2022,Chen2023}.
The main results are summarized below.

(1) Out of the six objects detected with the EVN, the radio emission in three objects is likely associated with an AGN driven wind, in two objects it is likely the coronal emission, and one object shows both a compact core component and a spatially separate extended component which are likely from a low-power jet.

(2) In the combined sample including our EVN and VLBA observations, 74\% (14/19) of the objects show either both radio and BLR winds or neither.
Of these objects, 86\% (12/14) are consistent with the interpretation that all of the wind objects are characterized by a high $L/L_{\rm Edd}$ ($\ge$ 0.66), while nearly all of the no wind objects are characterized by a low $L/L_{\rm Edd}$ ($\le$ 0.28).
This suggests that the AGN winds are probably driven by the radiation pressure.

(3) The wind indicators can be complicated by various aspects.
First, free-free absorption by AGN photoionized gas, if present, would flatten the radio spectral slope.
Second, a close to face-on view of a disk-like BLR could overestimate the $L/L_{\rm Edd}$.
The inclination may further affect a mild equatorial BLR wind, which may be weakened in a face-on view or enhanced in an edge-on view, in the intermediate $L/L_{\rm Edd}$ (0.2--0.4) objects.
Last, the radio outflow in the low $L/L_{\rm Edd}$ objects is possibly from a low-power jet, instead of a radiation pressure driven wind, as suggested by additional evidence.

Future studies on the connection between the radio outflow and the [O\,III] excess blue wing emission will help to clarify whether the pc-scale radio wind is connected to the NLR wind, if they occur on a comparable scale, as suggested in lower resolution radio observations \citep{Zakamska2014,Zakamska2016}.
The combination of integral field spectrograph and radio imaging on kpc scales could be beneficial to explore the interaction between the radio outflow (a wind or a jet) and the ambient medium.
New high S/N spectropolarimetry can further examine the inclination bias.
Finally, these results have to be taken with caution given the small sample.
A large sample is necessary to confirm these findings and draw a more reliable picture of the outflow multi-wavelength properties in RQ AGN.

\begin{acknowledgments}

We thank the anonymous referee for suggestions leading to the improvement of this work.
S.C. is supported in part at the Technion by a fellowship from the Lady Davis Foundation.
A.L. acknowledges support by the Israel Science Foundation (grant no.1008/18).
E.B. acknowledges support by a Center of Excellence of the Israel Science Foundation (grant no.2752/19).
The research leading to these results has received funding from the European Union's Horizon 2020 Research and Innovation Programme under grant agreement No.101004719 (OPTICON RadioNet Pilot).
The European VLBI Network is a joint facility of independent European, African, Asian, and North American radio astronomy institutes. Scientific results from data presented in this publication are derived from the following EVN project code(s): EC088.

\end{acknowledgments}

\vspace{5mm}
\facilities{EVN, e-MERLIN}

\software{AIPS \citep{Greisen2003}, DIFMAP \citep{Shepherd1994}, astropy \citep{Astropy2022}}

\bibliography{main.bbl}
\bibliographystyle{aasjournal}


\begin{table*}[ht!]
\caption{The EVN coordinates and their separations from the {\it Gaia} positions, and the beam and deconvolved source sizes in the EVN observations.}
\label{size}
\centering
\footnotesize
\begin{tabular}{ccccccccccc}
\hline\hline
Name & Frequency & \multicolumn{2}{c}{Coordinates} & Separation & \multicolumn{3}{c}{Beam size} & \multicolumn{3}{c}{Deconvolved source size} \\
& $\nu$ & R.A. & Dec. & $\Delta$ & $\theta_{\rm maj}$ & $\theta_{\rm min}$ & PA & $\theta_{\rm maj}$ & $\theta_{\rm min}$ & PA \\
& (GHz) & (hh:mm:ss) & (dd:mm:ss) & (mas) & (mas) & (mas) & (degree) & (mas) & (mas) & (degree) \\
(1) & (2) & (3) & (4) & (5) & (6) & (7) & (8) & (9) & (10) & (11) \\
\hline
\multirow{2}{*}{PG\,0049+171} & 4.9 & 00:51:54.7634 & +17:25:58.5085 & 0.4 & 6.08 & 2.07 & 9.3 & 3.08 & 1.09 & 5.1 \\
& 1.7 & 00:51:54.7633 & +17:25:58.5032 & 5.6 & 9.97 & 3.12 & 14.4 & $<$ 4.99 & 1.79 & \\
\hline
\multirow{2}{*}{PG\,1116+215} & 4.9 & 11:19:08.6783 & +21:19:17.9865 & 4.2 & 8.56 & 5.54 & 99.9 & 9.64 & 6.13 & 90.8 \\
& 1.7 & 11:19:08.6790 & +21:19:17.9872 & 5.1 & 27.6 & 9.59 & 6.4 & 19.64 & 14.06 & 13.7 \\
\hline
\multirow{2}{*}{PG\,1244+026} & 4.9 & 12:46:35.2530 & +02:22:08.7785 & 1.9 & 9.37 & 3.50 & 178.1 & $<$ 4.69 & $<$ 1.75 & \\
& 1.7 & 12:46:35.2529 & +02:22:08.7795 & 1.3 & 4.82 & 2.54 & 86.3 & 3.27 & $<$ 1.27 & \\
\hline
\multirow{2}{*}{PG\,1404+226} & 4.9 & 14:06:21.8901 & +22:23:46.5142 & 0.6 & 5.97 & 1.87 & 7.0 & 6.03 & 2.58 & 13.8 \\
& 1.7 & & & & & & & & & \\
\hline
\multirow{2}{*}{PG\,1416$-$129} & 4.9 & 14:19:03.8172 & $-$13:10:44.7860 & 0.3 & 8.34 & 2.61 & 4.4 & $<$ 4.17 & $<$ 1.31 & \\
& 1.7 & 14:19:03.8172 & $-$13:10:44.7910 & 5.2 & 5.55 & 3.99 & 57.2 & 2.88 & 2.37 & 85.6 \\
\hline
\multirow{3}{*}{PG\,1426+015} & 4.9 C1 & 14:29:06.5721 & +01:17:06.1521 & 2.3 & 7.84 & 2.63 & 4.9 & $<$ 3.92 & $<$ 1.32 & \\
& 4.9 C2 & 14:29:06.5728 & +01:17:06.1158 & 40.4 & 7.84 & 2.63 & 4.9 & 4.66 & $<$ 1.32 & \\
& 1.7 & & & & & & & & & \\
\hline
\end{tabular}
\begin{flushleft}
\vspace{-0.3cm}
\tablecomments{Columns:
(1) the name,
(2) the frequency,
(3) the right ascension of the centroid of EVN emission determined using JMFIT,
(4) the declination of the centroid of EVN emission determined using JMFIT,
(5) the separation between the EVN and the {\it Gaia} positions,
(6) the major axis of the beam in unit of mas,
(7) the minor axis of the beam in unit of mas,
(8) the position angle of the beam in unit of degree,
(9) the deconvolved major axis of the source in unit of mas,
(10) the deconvolved minor axis of the source in unit of mas,
(11) the deconvolved position angle of the source in unit of degree.}
\end{flushleft}
\end{table*}

\begin{turnpage}
\begin{table*}[ht!]
\caption{The $uv$ coverage, the core and total flux densities, and the background noise at 1.7 and 4.9 GHz in the full-array maps and the tapered maps, and the brightness temperature in the EVN observations.}
\label{flux}
\centering
\footnotesize
\begin{tabular}{ccccccccccc}
\hline\hline
Name & Frequency & \multicolumn{4}{c}{Full-array maps} & \multicolumn{4}{c}{Tapered maps} & $\log T_{\rm B}$ \\
& $\nu$ & $uv$-range & $S_{\rm total}$ & $S_{\rm core}$ & RMS & $uv$-range & $S_{\rm total}$ & $S_{\rm core}$ & RMS & \\
& (GHz) & (M$\lambda$) & (mJy) & (mJy\,beam$^{-1}$) & (mJy\,beam$^{-1}$) & (M$\lambda$) & (mJy) & (mJy\,beam$^{-1}$) & (mJy\,beam$^{-1}$) & (K) \\
(1) & (2) & (3) & (4) & (5) & (6) & (7) & (8) & (9) & (10) & (11) \\
\hline
\multirow{2}{*}{PG\,0049+171} & 4.9 & 3--170 & 0.29 $\pm$ 0.04 & 0.23 $\pm$ 0.02 & 0.020 & \multirow{2}{*}{3--60} & 0.29 $\pm$ 0.05 & 0.24 $\pm$ 0.02 & 0.024 & 6.84 \\
& 1.7 & 0--60 & 0.20 $\pm$ 0.03 & 0.16 $\pm$ 0.01 & 0.015 & & 0.25 $\pm$ 0.05 & 0.14 $\pm$ 0.02 & 0.022 & \\
\hline
\multirow{2}{*}{PG\,1116+215} & 4.9 & 2--170 & & $<$ 0.12 & 0.023 & \multirow{2}{*}{2--40} & 0.37 $\pm$ 0.08 & 0.16 $\pm$ 0.03 & 0.027 & 5.74 \\
& 1.7 & 0--60 & & $<$ 0.39 & 0.078 & & 1.33 $\pm$ 0.44 & 0.61 $\pm$ 0.14 & 0.116 & \\
\hline
\multirow{2}{*}{PG\,1244+026} & 4.9 & 2--170 & $<$ 0.19 $\pm$ 0.03 & 0.19 $\pm$ 0.03 & 0.027 & \multirow{2}{*}{2--40} & $<$ 0.19 $\pm$ 0.03 & 0.19 $\pm$ 0.03 & 0.028 & $>$ 6.26 \\
& 1.7 & 0--60 & 0.35 $\pm$ 0.05 & 0.21 $\pm$ 0.02 & 0.024 & & 0.33 $\pm$ 0.11 & 0.28 $\pm$ 0.06 & 0.049 & \\
\hline
\multirow{2}{*}{PG\,1404+226} & 4.9 & 1--170 & 0.44 $\pm$ 0.06 & 0.18 $\pm$ 0.02 & 0.021 & \multirow{2}{*}{1--50} & 0.47 $\pm$ 0.06 & 0.27 $\pm$ 0.02 & 0.025 & 6.37 \\
& 1.7 & 0--60 & & $<$ 0.24 & 0.048 & & & $<$ 0.25 & 0.049 & \\
\hline
\multirow{2}{*}{PG\,1416$-$129} & 4.9 & 2--160 & $<$ 0.39 $\pm$ 0.03 & 0.39 $\pm$ 0.03 & 0.032 & \multirow{2}{*}{2--50} & $<$ 0.38 $\pm$ 0.03 & 0.38 $\pm$ 0.03 & 0.035 & $>$ 6.78 \\
& 1.7 & 0--50 & 0.69 $\pm$ 0.11 & 0.53 $\pm$ 0.05 & 0.044 & & 0.72 $\pm$ 0.08 & 0.47 $\pm$ 0.03 & 0.036 & \\
\hline
\multirow{3}{*}{PG\,1426+015} & 4.9 C1 & 3--170 & $<$ 0.19 $\pm$ 0.04 & 0.19 $\pm$ 0.04 & 0.033 & \multirow{3}{*}{3--50} & $<$ 0.15 $\pm$ 0.04 & 0.15 $\pm$ 0.04 & 0.038 & $>$ 6.47 \\
& 4.9 C2 & 3--170 & 0.29 $\pm$ 0.10 & 0.15 $\pm$ 0.04 & 0.033 & & 0.30 $\pm$ 0.08 & 0.28 $\pm$ 0.04 & 0.038 & $>$ 6.59 \\
& 1.7 & 0--50 & & $<$ 0.23 & 0.046 & & & $<$ 0.24 & 0.047 & \\
\hline
\multirow{2}{*}{PG\,1012+008} & 4.9 & 1--170 & & $<$ 0.14 & 0.027 & & & & & \\
& 1.7 & 0--60 & & $<$ 0.15 & 0.030 & & & & & \\
\hline
\multirow{2}{*}{PG\,1211+143} & 4.9 & 3--170 & & $<$ 0.13 & 0.026 & & & & & \\
& 1.7 & 0--60 & & $<$ 0.27 & 0.053 & & & & & \\
\hline
\multirow{2}{*}{PG\,1626+554} & 4.9 & 2--140 & & $<$ 0.11 & 0.021 & & & & & \\
& 1.7 & 0--13 & & $<$ 0.10 & 0.020 & & & & & \\
\hline
\multirow{2}{*}{PG\,2112+059} & 4.9 & 3--170 & & $<$ 0.10 & 0.020 & & & & & \\
& 1.7 & 0--60 & & $<$ 0.16 & 0.031 & & & & & \\
\hline
\end{tabular}
\begin{flushleft}
\vspace{-0.3cm}
\tablecomments{Columns:
(1) the name,
(2) the frequency,
(3) the $uv$-range of the full-array map,
(4) the total flux density of the full-array map,
(5) the core flux density of the full-array map,
(6) the background noise of the full-array map,
(7) the $uv$-range of the tapered map,
(8) the total flux density of the tapered map,
(9) the core flux density of the tapered map,
(10) the background noise of the tapered map,
(11) the brightness temperature in logarithm scale.}
\end{flushleft}
\end{table*}
\end{turnpage}

\begin{table*}[ht!]
\caption{The BH mass, the VLA 5~GHz flux, the {\it XMM-Newton} 0.2--12~keV flux, and the luminosity ratio of the EVN/VLBA at 5~GHz to the {\it XMM-Newton} at 0.2--12~keV of the objects.}
\label{ratio}
\centering
\footnotesize
\begin{tabular}{cccccccc}
\hline\hline
Name & Array & $z$ & Scale & $\log M_{\rm BH}/M_{\odot}$ & $S_{\rm VLA}$ & $f_{\rm XMM-Newton}$ & $\log L_{\rm R}/L_{\rm X}$ \\
& & & (pc\,mas$^{-1}$) & & (mJy) & ($10^{-12}$\,erg\,s$^{-1}$\,cm$^{-2}$) & \\
(1) & (2) & (3) & (4) & (5) & (6) & (7) & (8) \\
\hline
PG\,0049+171 & \multirow{10}{*}{EVN} & 0.065 & 1.26 & 7.73 & 0.66$^{a}$ & 39.10 $\pm$ 0.29 & $-6.54 \pm 0.04$ \\
PG\,1116+215 & & 0.176 & 3.08 & 8.35 & 1.94$^{a}$ & 11.10 $\pm$ 0.05 & $-6.15 \pm 0.08$ \\
PG\,1244+026 & & 0.048 & 1.00 & 6.15 & 0.47$^{a}$ & 12.50 $\pm$ 0.03 & $-6.13 \pm 0.07$ \\
PG\,1404+226 & & 0.098 & 1.89 & 6.52 & 0.89$^{a}$ & 0.88 $\pm$ 0.01 & $-5.00 \pm 0.05$ \\
PG\,1416$-$129 & & 0.129 & 2.39 & 8.74 & 0.80$^{a}$ & 5.31 $\pm$ 0.04 & $-5.44 \pm 0.03$ \\
PG\,1426+015 & & 0.087 & 1.69 & 8.67 & 0.93$^{a}$ & 18.50 $\pm$ 0.08 & $-6.30 \pm 0.09$ \\
PG\,1012+008 & & 0.187 & 3.24 & 8.01 & 0.74$^{a}$ & & \\
PG\,1211+143 & & 0.081 & 1.59 & 7.64 & 1.17$^{a}$ & 11.00 $\pm$ 0.03 & $< -6.24$ \\
PG\,1626+554 & & 0.134 & 2.45 & 8.13 & 0.32$^{a}$ & 7.78 $\pm$ 0.06 & $< -6.16$ \\
PG\,2112+059 & & 0.459 & 5.99 & 8.85 & 0.76$^{a}$ & 0.22 $\pm$ 0.01 & $< -4.65$ \\
\hline
PG0026+129 & \multirow{17}{*}{VLBA} & 0.142 & 3.26 & 7.74 & 0.20$^{a}$ & 9.25 $\pm$ 0.10 & $-5.88 \pm 0.04$ \\
PG0050+124 & & 0.060 & 1.17 & 6.99 & 2.41$^{b}$ & 14.90 $\pm$ 0.04 & $-6.10 \pm 0.09$ \\
PG0052+251 & & 0.155 & 2.71 & 8.64 & 0.68$^{b}$ & 13.70 $\pm$ 0.06 & $-6.06 \pm 0.03$ \\
PG0157+001 & & 0.164 & 3.81 & 8.00 & 5.58$^{a}$ & 2.82 $\pm$ 0.05 & $-5.34 \pm 0.06$ \\
PG0921+525 & & 0.035 & 0.75 & 6.87 & 1.87$^{a}$ & 55.80 $\pm$ 0.10 & $-6.05 \pm 0.01$ \\
PG0923+129 & & 0.029 & 0.62 & 6.82 & 2.82$^{c}$ & 34.10 $\pm$ 0.13 & $-6.61 \pm 0.06$ \\
PG1149$-$110 & & 0.050 & 0.98 & 7.34 & 2.27$^{b}$ & 9.12 $\pm$ 0.08 & $-5.64 \pm 0.02$ \\
PG1216+069 & & 0.334 & 8.52 & 9.06 & 4.95$^{a}$ & 3.65 $\pm$ 0.03 & $-4.04 \pm 0.01$ \\
PG1351+640 & & 0.087 & 1.92 & 8.49 & 20.0$^{a}$ & 0.95 $\pm$ 0.02 & $-3.93 \pm 0.01$ \\
PG1501+106 & & 0.036 & 0.77 & 8.11 & 0.50$^{a}$ & 32.90 $\pm$ 0.13 & $-6.31 \pm 0.04$ \\
PG1534+580 & & 0.030 & 0.64 & 7.71 & 1.80$^{a}$ & 17.10 $\pm$ 0.10 & $-6.41 \pm 0.08$ \\
PG1612+261 & & 0.131 & 2.35 & 7.69 & 5.58$^{b}$ & 8.07 $\pm$ 0.09 & $< -6.31$ \\
PG2304+042 & & 0.042 & 0.83 & 7.91 & 0.77$^{b}$ & 30.00 $\pm$ 0.13 & $-6.35 \pm 0.03$ \\
PG1351+236 & & 0.055 & 1.19 & 8.10 & $<$ 0.25$^{a}$ & & \\
PG1440+356 & & 0.077 & 1.55 & 7.09 & 1.24$^{b}$ & $10.90 \pm 0.05$ & $< -6.44$ \\
PG1613+658 & & 0.139 & 2.25 & 8.89 & 3.03$^{b}$ & $10.70 \pm 0.19$ & $< -6.44$ \\
PG2130+099 & & 0.062 & 1.24 & 7.49 & 2.18$^{b}$ & $10.00 \pm 0.06$ & $< -6.46$ \\
\hline
\end{tabular}
\begin{flushleft}
\vspace{-0.3cm}
\tablecomments{Columns:
(1) the name,
(2) the observing array,
(3) the redshift,
(4) the physical scale,
(5) the BH mass in logarithm scale from \citet{Davis2011},
(6) the VLA flux density at 5~GHz with the A configuration from literature,
(7) the X-ray flux at 0.2--12.0~keV in units of 10$^{-12}$\,erg\,s$^{-1}$\,cm$^{-2}$ from the {\it XMM-Newton} DR12 catalogue \citep{Webb2020},
(8) the ratio of the EVN/VLBA core luminosity at 5~GHz to the {\it XMM-Newton} luminosity at 0.2--12~keV in logarithm scale.
References:
a - \citet{Kellermann1989}, b - \citet{Alhosani2022}, c - \citet{Berton2018}.}
\end{flushleft}
\end{table*}

\begin{turnpage}
\begin{table*}[ht!]
\caption{The spectral slope, the compactness, the dominant radio origin, the C\,IV shift, whether the BLR wind is present, the Eddington ratio, and the grouping based on whether the radio and/or BLR winds are present in the combined sample with EVN and VLBA observations.}
\label{result}
\centering
\footnotesize
\begin{tabular}{ccccccccccc}
\hline\hline
Name & Array & $\alpha_{\rm EVN/VLBA}$ & $\alpha_{\rm VLA}$ & $\frac{S_{\rm core}}{S_{\rm total}}$ & $\frac{S_{\rm EVN/VLBA}}{S_{\rm VLA}}$ & Radio origin & C\,IV $\frac{v_{\rm shift}}{\rm{FWHM}}$ & BLR wind & $\log L/L_{\rm Edd}$ & Group \\
(1) & (2) & (3) & (4) & (5) & (6) & (7) & (8) & (9) & (10) & (11) \\
\hline
PG\,0049+171 & \multirow{8}{*}{EVN} & 0.13 $\pm$ 0.23 & $-0.30^{a}$ & 0.8 & 0.4 & Corona & 0.021 & No & $-$1.15 & No wind \\
PG\,1116+215 & & $-$1.18 $\pm$ 0.36 & $-0.59^{a}$ & 0.4 & 0.2 & Wind & $-$0.108 & Yes & $-$0.18 & Radio + BLR winds \\
PG\,1244+026 & & $-$0.52 $\pm$ 0.33 & $-0.87^{b}$ & 1.0 & 0.4 & Wind & $-$0.175 & Yes & 0.49 & Radio + BLR winds \\
PG\,1404+226 & & $>$ 0.09 $\pm$ 0.08 & $-0.63^{a}$ & 0.4 & 0.5 & Wind & $-$0.538 & Yes & 0.59 & Radio + BLR winds \\
PG\,1416$-$129 & & $-$0.59 $\pm$ 0.13 & $0.11^{c}$ & 1.0 & 0.5 & Corona & 0.001 & No & $-$1.02 & No wind \\
PG\,1426+015 & & $>$ $-$0.07 $\pm$ 0.13 & $-0.18^{a}$ & 0.4 & 0.5 & Jet? & $-$0.021 & No & $-$0.93 & No wind \\
C1 & & $>$ $-$0.40 $\pm$ 0.26 & & 1.0 & & & & & \\
C2 & & $>$ 0.17 $\pm$ 0.14 & & 0.5 & & & & & \\
\hline
PG\,0026+129 & \multirow{13}{*}{VLBA} & $>$ 0.89 $\pm$ 0.14 & $-0.31^{d}$ & 0.7 & 1.6 & Corona & 0.075 & No & 0.30 & No wind \\
PG\,0050+124 & & $-$0.90 $\pm$ 0.15 & $-1.45^{a}$ & 0.4 & 0.2 & Wind & $-$0.040 & Yes & 0.03 & Radio + BLR winds \\
PG\,0052+251 & & $-$0.13 $\pm$ 0.16 & $0.93^{a}$ & 0.8 & 0.4 & Corona & $-$0.019 & No & $-$0.68 & No wind \\
PG\,0157+001 & & $-$0.90 $\pm$ 0.16 & $-0.60^{a}$ & 0.1 & 0.3 & Wind & $-$0.344 & Yes & $-$0.18 & Radio + BLR winds \\
PG\,0921+525 & & $<$ $-$0.12 $\pm$ 0.07 & $-0.17^{a}$ & 0.7 & 0.9 & Wind & 0.166 & No & $-$0.51 & Radio wind only \\
PG\,0923+129 & & $<$ $-$1.40 $\pm$ 0.09 & $-0.94^{e}$ & 0.3 & 0.1 & Wind & 0.112 & No & $-$0.40 & Radio wind only \\
PG\,1149$-$110 & & $-$0.31 $\pm$ 0.12 & $0.48^{a}$ & 0.7 & 0.3 & Corona & $-$0.123 & Yes & $-$0.70 & BLR wind only \\
PG\,1216+069 & & 2.18 $\pm$ 0.10 & $0.52^{a}$ & 1.0 & 1.4 & Corona & 0.180 & No & $-$0.56 & No wind \\
PG\,1351+640 & & $-$0.43 $\pm$ 0.01 & $-0.64^{a}$ & 0.4 & 0.3 & Jet & 0.052 & No & $-$1.29 & No wind \\
PG\,1501+106 & & 0.08 $\pm$ 0.21 & $0.17^{a}$ & 0.3 & 1.3 & Corona & $-$0.058 & No & $-$1.32 & No wind \\
PG\,1534+580 & & $<$ $-$0.70 $\pm$ 0.17 & $-0.73^{f}$ & 0.6 & 0.1 & Wind? & 0.015 & No & $-$1.33 & Radio wind only \\
PG\,1612+261 & & $<$ $-$1.61 $\pm$ 0.11 & $-1.57^{a}$ & & $<$ 0.01 & Wind & 0.114 & No & $-$0.41 & Radio wind only \\
PG\,2304+042 & & $-$0.09 $\pm$ 0.12 & $0.67^{a}$ & 0.6 & 0.6 & Corona & $-$0.044 & No & $-$1.52 & No wind \\
\hline
\end{tabular}
\begin{flushleft}
\vspace{-0.3cm}
\tablecomments{Columns:
(1) the name,
(2) the observing array,
(3) the EVN/VLBA spectral slope of the total flux density at 1.7--4.9 GHz,
(4) the VLA spectral slope from literature,
(5) the EVN/VLBA core to total flux ratio at 5~GHz,
(6) the ratio of EVN/VLBA total flux to VLA A-configuration core flux at 5~GHz,
(7) the origin of the radio emission,
(8) the C\,IV line velocity shift compared to H$\beta$ line in units of C\,IV FWHM from \citet{Baskin2005},
(9) whether the BLR wind is present,
(10) the Eddington ratio in logarithm scale from \citet{Davis2011},
(11) the grouping based on whether the radio and/or BLR winds are present.
References:
a - \citet{Laor2019} at 5--8.5~GHz,
b - \citet{Berton2018} at 1.5--5~GHz,
c - \citet{Barvainis1996} at 5--8.5~GHz,
d - \citet{Baldi2022} at 5--8.5~GHz,
e - \citet{Schmitt2001a,Berton2018} at 5--8.5~GHz,
f - the FIRST \citep{Helfand2015} and \citet{Leipski2006} at 1.5--5~GHz.}
\end{flushleft}
\end{table*}
\end{turnpage}

\end{document}